\documentclass[11pt,twoside,openright,notitlepage]{article}
\usepackage{amssymb}
\usepackage{amsfonts}
\usepackage{float}
\usepackage{makeidx}
\usepackage{graphicx}
\usepackage{amsmath}

\input{tcilatex}
\begin{document}

\title{{\Large HYDRODYNAMIC COUPLED MODES OF A NEMATIC UNDER A TEMPERATURE
GRADIENT AND A UNIFORM GRAVITATIONAL FIELD}}
\author{J. Camacho$^{1}$ and R. F. Rodr\'{\i}guez$^{2,3,4}$. \\
%EndAName
$^{1}$Maestr\'{\i}a en Ciencias de la Complejidad, \\
Universidad Aut\'{o}noma de la Ciudad de M\'{e}xico.\\
$^{2}$Instituto de F\'{\i}sica, Universidad Nacional Aut\'{o}noma de M\'{e}%
xico. \\
$^{3}$Correspondence author. E-mail: zepeda@fisica.unam.mx.\\
$^{4}$Fellow of SNI, M\'{e}xico.}
\maketitle
\date{ }

\begin{abstract}
{\small Fluctuating hydrodynamics $(FH)$ describes the dynamics of the
fluctuations for fluids at mesoscopic scales. Here we use this approach to
study the fluctuations of the hydrodynamic variables of a thermotropic
nematic liquid crystal $\ (NLC)\ $in a nonequilibrium steady state $(NESS)$.
This state is induced by an externally imposed temperature gradient and a
uniform gravity field. We calculate analytically both, the equilibrium and
nonequilibrium hydrodynamic modes. We find that in this $NESS$\ the
nonequilibrium effects produced by the external gradients only affect the
longitudinal variables. This gives rise to a pair of sound modes, one
orientation mode of the director and two visco-heat modes formed by the
coupling of the shear and thermal modes. We also find that the last three
modes exhibit the largest changes.\emph{\ }The analytical expressions that
we have found for the visco-heat modes imply that the heat and shear modes
of the $NLC$ are coupled, that they reduce to those of simple fluid in the
isotropic limit and that these modes may become propagative, a feature that
also occurs in the simple fluid. In the isotropic limit of the nematic our
results also reduce to the hydrodynamic modes of a simple fluid in the
presence of the same temperature gradient and the pressure gradient produced
by the gravity field.}\bigskip \medskip
\end{abstract}

\section{INTRODUCTION}

The Landau and Lifshitz theory of hydrodynamic fluctuations close to
equilibrium \cite{landau1}, was put on a firm basis within the framework of
the general theory of stationary Gaussian Markov processes by Fox and
Uhlenbeck \cite{fox1}, \cite{fox2}. In fluctuating hydrodynamics ($FH$)\ the
usual deterministic hydrodynamic equations are supplemented with random
dissipative fluxes of thermal origin, obeying fluctuation-dissipation
relations. This approach has matched the theory of Onsager and Machlup with
the approach of Landau and Lifshitz, for systems where the basic state
variables do not posses a definite time reversal symmetry, leading to
Langevin-like stochastic equations for the evolution of the fluctuations of
the state variables \cite{onsager1}, \cite{onsager1a}, \cite{onsager2}. In
this way fluctuating hydrodynamics provides a systematic method for
assessing the nature of spontaneous fluctuations induced by intrinsic
thermal noise.

The Fox and Uhlenbeck's scheme has been applied to simple fluids and their
binary mixtures \cite{foch}, \cite{cohen}, \cite{ortiz}; however, more
recently it has been also verified that $FH$ can be extended to deal with
thermally excited fluctuations in complex fluids in stationary
nonequilibrium states \cite{lopezdeharo}, \cite{rodriguez0}, \cite%
{rodriguez00}, \cite{rodriguez2}. In spite of the fact that the theory of
fluctuations in nonequilibrium fluids was initiated in the late 70's \cite%
{kirkpatrick1}, \cite{tremblay1}, and pursued by many authors \cite{fuller}, 
\cite{rubi2}, still nowadays several questions concerning the nature of
hydrodynamic fluctuations in $NESS$ are of current active interest. One of
these issues is the long-range character of these fluctuations even far away
from instability points. It has been shown theoretically that the existence
of the so called generic scale invariance is the origin of the long range
nature of the correlation functions \cite{dorfman}. However, in spite of the
considerable interest in fluctuations about dissipative steady states of
simple fluids during the last two decades, there are few similar studies for
equilibrium or nonequilibrium stationary states of complex fluids \cite%
{sengers2}, \cite{deutch}.

The basic purpose of the present paper is to describe the dynamics of the
fluctuations of the hydrodynamic variables of a nematic liquid crystals in a 
$NESS$ induced by a stationary temperature gradient and under the influence
of gravity \cite{tesis}. More specifically, we compare the fluctuations in
the presence of a dissipative thermodynamic force like a stationary
temperature gradient, with those in the presence of a conservative
thermodynamic force like gravity. This comparison has been analyzed for a
simple fluid \cite{segre}, and since gravity is a small force, significant
changes in the fluctuations only occur at wave numbers that are too small to
be probed experimentally. However, it is important to know theoretically how
such small force can affect the fluctuations also in a liquid crystal. To
our knowledge, this issue has only been explored for liquid crystals in the
Refs. \cite{miyakawa}, \cite{migranov} for the same $NESS$ considered in
this work. However, we consider that their conclusions are not definitive
because in the isotropic limit, they do not reduce to the well known
corresponding visco-heat modes of a simple fluid \cite{lekkerkerker-boon}.
References \cite{miyakawa}, \cite{migranov} predict that the thermal and
director diffusive modes are coupled. In contrast, the analytical
expressions that we have found for the visco-heat modes imply, on the one
hand, that the heat and shear modes of the $NLC$ are coupled and that they
reduce to those of simple fluid in the isotropic limit.\emph{\ }Furthermore,%
\emph{\ }our expressions\emph{\ }also predict that these modes may become
propagative, a feature that also occurs in the simple fluid \cite{segre}.

To this end and on the basis of $FH$, we first evaluate the equilibrium and
nonequilibrium hydrodynamic modes of the $NLC$ and then we show that among
the longitudinal modes, which are the only ones affected by the external
gradients, there is a pair of visco-heat modes constituted by the coupling
of the shear and thermal modes. These modes are more general than those
reported so far in the literature for a $NLC$ in a $NESS$ produced by a
temperature gradient \cite{miyakawa}, \cite{migranov}, because they reduce
to the corresponding expressions in the isotropic limit of the $NLC$\ \cite%
{buka}, \cite{Rusos}, \cite{Leslie}, to those of a simple fluid \cite{ortiz}%
, \cite{segre}, \cite{lekkerkerker-boon}.

\section{MODEL}

Consider a quiscent thermotropic nematic liquid crystal thin layer of
thickness $d$ confined between two parallel plates in a homeotropic
arrangement, $\widehat{n}_{0}=\left( 0,0,1\right) $. The $NLC$\ is in the
presence of a uniform gravitational field $\overrightarrow{g}=-g\widehat{z}$%
, where $\widehat{z}$ denotes the unitary vector along the $z$-axis as
depicted in Fig. 1. The transverse dimensions of the cell along the $x$ and $%
y$ directions are large compared to $d$.

\FRAME{ftbphFU}{3.3956in}{2.0174in}{0pt}{\Qcb{Schematic representation of
the homeotropic nematic cell under the influence of a constant gravitational
field $\protect\overrightarrow{g}$ and an external, uniform, temperature
gradient $\protect\nabla T$. }}{}{fig6-1art.wmf}{\special{language
"Scientific Word";type "GRAPHIC";maintain-aspect-ratio TRUE;display
"USEDEF";valid_file "F";width 3.3956in;height 2.0174in;depth
0pt;original-width 9.103in;original-height 5.3947in;cropleft "0";croptop
"1";cropright "1";cropbottom "0";filename
'../../AppData/Local/Temp/Rar$DIa0.729/fig6-1art.wmf';file-properties
"NPEU";}}

The plates are maintained at the uniform temperatures $T_{1}$ $>$ $T_{2}$ so
that a constant temperature gradient 
\begin{equation}
\nabla _{z}T(z)=-\alpha \widehat{z}  \label{2-1}
\end{equation}%
is established between them. The presence of the gravitational field induces
a constant pressure gradient $\nabla _{z}p$ given by 
\begin{equation}
\nabla _{z}p(z)=-\rho g\widehat{z},  \label{2-2}
\end{equation}%
where $\rho $\ is the mass density. Furthermore, if the temperature
difference between the plates is of only a few degrees, we may assume that $%
\nabla _{z}T$ and $\nabla _{z}p$\ do not generate flows or hydrodynamic
instabilities, so that the $NLC$ is in a quiscent $NESS$. This state is
described by $\psi ^{st}\equiv \left\{ v_{i}^{st}\text{,}\rho ^{st}(z)\text{,%
}s^{st}(z)\text{,}\widehat{n}^{st}(z)\right\} $, where $v_{i}^{st}=0$ is the
hydrodynamic velocity and $\rho ^{st}$, $s^{st}$, $\widehat{n}^{st}$ denote,
respectively, the mass, entropy, $s^{st}(z)$ and director $\widehat{n}%
^{st}(z)$ local densities of the $NLC$ in the stationary state. The gradient
of the state variables $\psi ^{st}$ can be expanded in a Taylor series
around the equilibrium state $(p_{0},T_{0})$ in terms of the external
gradients. For the boundary conditions $T(z=-d/2)=T_{1}$, $T(z=d/2)=T_{2}$
and to first order in these gradients, the stationary temperature profile is
then given by 
\begin{equation}
T^{st}(z)=T_{0}\left( 1-\frac{\alpha }{T_{0}}z\right) ,  \label{2-3}
\end{equation}%
where $T_{0}\equiv T^{st}\left( z=0\right) =\left( T_{1}+T_{2}\right) /d$ , $%
\alpha \equiv \left( T_{1}-T_{2}\right) /d$ and 
\begin{equation}
\nabla _{z}s^{st}=\frac{c_{p}}{T_{0}}X\widehat{z},  \label{2-4}
\end{equation}

\begin{equation}
\nabla _{z}\rho ^{st}=-\rho _{0}\beta \left( X+\frac{g\beta T_{0}}{%
c_{p}\left( \gamma -1\right) }\right) \widehat{z}.  \label{2-5}
\end{equation}%
Note that the effective temperature gradient $X\equiv -\alpha +\frac{g\beta
T_{0}}{c_{p}}$ contains the contributions of both external gradients ($%
\alpha $ and $g$), and $c_{p}$, $\beta $, $c_{T}$ are, respectively, the
specific heat at constant pressure, the thermal expansion coefficient and
the isothermal sound velocity of the nematic. To arrive at Eqs. (\ref{2-4})
and (\ref{2-5}) we have used the thermodynamic relations $\beta ^{2}\equiv
\left( \gamma -1\right) c_{p}/T_{0}c_{s}^{2}$ and $\gamma
=c_{s}^{2}/c_{T}^{2}$, where $c_{s}$ is the adiabatic sound velocity.

\subsection{Fluctuating nematodynamics}

The mass conservation equation for $\rho $, the equation of motion for $%
\overrightarrow{v}$,\ the balance equation for $s$ and the relaxation
equation for the director field $\widehat{n}$ are given, respectively, by 
\cite{buka}, \cite{de gennes}, \cite{Chandra}, \cite{Landau2}, \cite{camacho}%
,

\begin{equation}
\left( \frac{\partial }{\partial t}+v_{j}\nabla _{j}\right) \rho +\rho
\nabla _{l}v_{l}=0,  \label{3-1}
\end{equation}%
\begin{equation}
\rho \left( \frac{\partial }{\partial t}+v_{j}\nabla _{j}\right)
v_{i}=-\nabla _{i}p+\nabla _{j}\sigma _{ij}^{\prime }-\nabla _{j}\left( \Phi
_{jl}\nabla _{i}n_{l}\right) -\frac{1}{2}\nabla _{j}\left( \lambda
_{imj}h_{m}\right) +\rho f_{i},  \label{3-2}
\end{equation}%
\begin{equation}
\rho T\left( \frac{\partial s}{\partial t}+v_{i}\nabla _{i}s\right) =\sigma
_{ij}^{\prime }\nabla _{j}v_{i}-\nabla _{j}q_{j}+h_{k}\mathcal{N}_{k}
\label{3-3}
\end{equation}%
\begin{equation}
\frac{\partial n_{i}}{\partial t}+v_{j}\nabla _{j}n_{i}=\frac{1}{2}\lambda
_{ijk}\nabla _{j}v_{k}+\mathcal{N}_{i},  \label{3-4}
\end{equation}%
where $p\left( \overrightarrow{r},t\right) $ is the pressure field and the
tensor $\sigma _{ij}^{\prime }$ denotes the momentum current 
\begin{equation}
\sigma _{ij}^{\prime }\equiv \nu _{ijlm}\nabla _{m}v_{l}.  \label{3-5}
\end{equation}%
The viscous tensor $\nu _{ijkl}$ is 
\begin{gather}
\nu _{ijkl}\equiv \nu _{2}(\delta _{jl}\delta _{ik}+\delta _{il}\delta
_{jk})+2(\nu _{1}+\nu _{2}-2\nu _{3})n_{i}n_{k}n_{l}  \notag \\
+(\nu _{3}-\nu _{2})(n_{j}n_{l}\delta _{ik}+n_{j}n_{k}\delta
_{il}+n_{i}n_{k}\delta _{jl}+n_{i}n_{l}\delta _{jk})  \notag \\
+(\nu _{4}-\nu _{2})\delta _{ij}\delta _{kl}+(\nu _{5}-\nu _{4}+\nu
_{2})(\delta _{ij}n_{k}n_{l}+\delta _{kl}n_{i}n_{j})  \label{3-6}
\end{gather}%
where the set $\nu _{i}=\left\{ \nu _{1},\nu _{2},\nu _{3},\gamma
_{1},\gamma _{2}\right\} $ denotes the five nematic viscosity coefficients
of a nematic in the notation of Harvard \cite{de gennes}; $f_{i}$ is the
total body force acting on the nematic; $\Phi _{ki}$ is given by%
\begin{equation}
\Phi _{ki}=K_{ikrj}\nabla _{j}n_{r},  \label{3-7}
\end{equation}%
where the fourth order tensor $K_{ijkl}$ depends on the elastic constants $%
K_{1}$ (splay), $K_{2}$ (twist), $K_{3}$ (bend) and it is defined in terms
of the Levi-Civitta tensor $\epsilon _{ijk}$ by 
\begin{equation}
K_{ijkl}=K_{1}\delta _{ij}\delta _{kl}+K_{2}n_{p}\epsilon
_{pij}n_{q}\epsilon _{qkl}+K_{3}n_{j}n_{l}\delta _{ik}.  \label{3-8}
\end{equation}%
\ The third order tensor in Eq. (\ref{3-4}) 
\begin{equation}
\lambda _{kji}\equiv (\lambda -1)\delta _{kj}^{\perp }n_{i}+(\lambda
+1)\delta _{ki}^{\perp }n_{j},  \label{3-9}
\end{equation}%
depends on the orientational viscosities through $\lambda \equiv -\gamma
_{1}/\gamma _{2}$. The vector $q_{l}$ is the heat flux 
\begin{equation}
q_{l}\equiv -\kappa _{lj}\nabla _{j}T,  \label{3-10}
\end{equation}%
where $\kappa _{ij}$ is the thermal conductivity tensor 
\begin{equation}
\kappa _{ij}=\kappa _{\perp }\delta _{ij}+\kappa _{a}n_{i}n_{j},
\label{3-11}
\end{equation}%
with anisotropy $\kappa _{a}\equiv \kappa _{\parallel }-\kappa _{\perp }$,
being $\kappa _{\perp }$ and $\kappa _{\parallel }$ its perpendicular and
parallel components with respect to the director field.\ $\mathcal{N}_{i}$
is the quasi-current associated with the director%
\begin{equation}
\mathcal{N}_{i}=\frac{1}{\gamma _{1}}\delta _{ik}^{\perp }h_{k}  \label{3-12}
\end{equation}%
and the molecular field $h_{i}$ is 
\begin{equation}
h_{i}=\delta _{ir}^{\perp }K_{rjkl}\nabla _{j}\nabla _{l}n_{k}-\delta
_{iq}^{\perp }\left( \frac{1}{2}\frac{\partial }{\partial n_{q}}K_{pjkl}-%
\frac{\partial }{\partial n_{p}}K_{qjkl}\right) \nabla _{j}n_{p}\nabla
_{l}n_{k}.  \label{3-13}
\end{equation}

It should be stressed that Eqs. (\ref{3-1})-(\ref{3-4}) are valid for any
motion of the nematic including the hydrodynamic deviations (fluctuations)\
of the state variables from the $NESS$. If only the linear deviations, $%
\delta \rho \left( \overrightarrow{r},t\right) =\rho \left( \overrightarrow{r%
},t\right) -\rho ^{st}$, $\delta v_{i}\left( \overrightarrow{r},t\right)
=v_{i}\left( \overrightarrow{r},t\right) $, $\delta s\left( \overrightarrow{r%
},t\right) =s\left( \overrightarrow{r},t\right) -s^{st}$, $\delta
n_{i}\left( \overrightarrow{r},t\right) =n_{i}\left( \overrightarrow{r}%
,t\right) -n_{i}^{st}$ around the stationary state $v_{i}^{st}=0$, $%
n_{i}^{st}=cte$, $h_{l}^{st}=0$ are considered and $\Phi _{lj}^{st}=0$, the
linearized equations associated with Eqs. (\ref{3-1})-(\ref{3-4}) read%
\begin{equation}
\frac{\partial }{\partial t}\delta \rho =-\delta v_{j}\nabla _{j}\rho
^{st}-\rho ^{st}\nabla _{l}\delta v_{l},  \label{3-14}
\end{equation}%
\begin{equation}
\rho ^{st}\frac{\partial }{\partial t}\delta v_{i}=-\nabla _{i}\delta p-%
\frac{1}{2}\lambda _{kji}^{st}\nabla _{j}\delta h_{k}+\nabla _{j}\delta
\sigma _{ij}^{\prime }-g\delta _{iz}\delta \rho ,  \label{3-15}
\end{equation}%
\begin{equation}
\rho ^{st}T^{st}\left( \frac{\partial }{\partial t}\delta s+\delta
v_{j}\nabla _{j}s^{st}\right) =-\nabla _{l}\delta q_{l},  \label{3-16}
\end{equation}%
\begin{equation}
\frac{\partial }{\partial t}\delta n_{i}=\frac{1}{2}\lambda
_{ijk}^{st}\nabla _{j}\delta v_{k}+\delta \mathcal{N}_{i},  \label{3-17}
\end{equation}%
with

\begin{equation}
\delta \sigma _{ij}^{\prime }\equiv \nu _{ijkl}^{st}\nabla _{l}\delta v_{k}.
\label{3-18}
\end{equation}%
\begin{equation}
\delta q_{l}\equiv -\delta \kappa _{lj}\nabla _{j}T^{st}-\kappa
_{lj}^{st}\nabla _{j}\delta T,  \label{3-19}
\end{equation}%
\begin{equation}
\delta \mathcal{N}_{i}\equiv \frac{1}{\gamma _{1}}\left( \delta _{ik}^{\perp
}\right) ^{st}\delta h_{k},  \label{3-20}
\end{equation}%
where%
\begin{equation}
\lambda _{kji}^{st}\equiv (\lambda -1)\left( \delta _{kj}^{\perp }\right)
^{st}n_{i}^{st}+(\lambda +1)\left( \delta _{ki}^{\perp }\right)
^{st}n_{j}^{st},  \label{3-21}
\end{equation}%
\begin{equation}
\delta h_{i}=\left( \delta _{ir}^{\perp }\right) ^{st}K_{rjkl}^{st}\nabla
_{j}\nabla _{l}\delta n_{k},  \label{3-22}
\end{equation}%
\begin{gather}
\nu _{ijkl}^{st}\equiv \nu _{2}(\delta _{jl}\delta _{ik}+\delta _{il}\delta
_{jk})+2(\nu _{1}+\nu _{2}-2\nu _{3})n_{i}^{st}n_{j}^{st}n_{k}^{st}n_{l}^{st}
\notag \\
+(\nu _{3}-\nu _{2})(n_{j}^{st}n_{l}^{st}\delta
_{ik}+n_{j}^{st}n_{k}^{st}\delta _{il}+n_{i}^{st}n_{k}^{st}\delta
_{jl}+n_{i}^{st}n_{l}^{st}\delta _{jk})  \notag \\
+(\nu _{4}-\nu _{2})\delta _{ij}\delta _{kl}+(\nu _{5}-\nu _{4}+\nu
_{2})(\delta _{ij}n_{k}^{st}n_{l}^{st}+\delta _{kl}n_{i}^{st}n_{j}^{st}),
\label{3-23}
\end{gather}%
\begin{equation}
\delta \kappa _{ij}=\kappa _{a}\left( n_{i}\delta n_{j}+\delta
n_{i}n_{j}\right) ,  \label{3-24}
\end{equation}%
\begin{equation}
\kappa _{ij}^{st}=\kappa _{\perp }\delta _{ij}+\kappa
_{a}n_{i}^{st}n_{j}^{st},  \label{3-25}
\end{equation}%
\begin{equation}
K_{ijkl}^{st}=K_{1}\delta _{ij}\delta _{kl}+K_{2}n_{p}\epsilon
_{pij}n_{q}\epsilon _{qkl}+K_{3}n_{j}n_{l}\delta _{ik},  \label{3-26}
\end{equation}%
where we have defined $\left( \delta _{ir}^{\perp }\right) ^{st}=\delta
_{ir}-n_{i}^{st}n_{r}^{st}$. In Eq. (\ref{3-15}) the total volumetric force
is $f_{i}=-\delta _{iz}g$.

Following Landau and Lifshitz \cite{landau1}, we now introduce fluctuating
components into the momentum current $\sigma _{ij}^{\prime }\left( \vec{r}%
,t\right) $, the heat flux $q_{l}$ and the relaxation quasi-current $%
\mathcal{N}_{i}$ of the orientation of the nematic. These stochastic
components are denoted, respectively, by $\nabla _{j}\Sigma _{ij}\left( \vec{%
r},t\right) $, $\pi _{i}\left( \vec{r},t\right) $, $\Upsilon _{i}\left( \vec{%
r},t\right) $, and are chosen as zero averaged stochastic processes%
\begin{equation}
\left\langle \Sigma _{ij}\left( \vec{r},t\right) \right\rangle =\left\langle
\pi _{i}\left( \vec{r},t\right) \right\rangle =\left\langle \Upsilon
_{i}\left( \vec{r},t\right) \right\rangle =0,  \label{3-27}
\end{equation}%
satisfying fluctuation-dissipation relations ($FDR$) which have the same
form as in equilibrium, but replacing the equilibrium temperature by $T^{st}$%
, \cite{buka}, \cite{rodriguez1}. These relations are%
\begin{equation}
\left\langle \Sigma _{\alpha j}(\overrightarrow{r},t)\Sigma _{\beta l}(%
\overrightarrow{r}^{\prime },t^{\prime })\right\rangle =2k_{B}T^{st}\nu
_{\alpha \beta jl}^{st}\delta \left( \overrightarrow{r}-\overrightarrow{r}%
^{\prime }\right) \delta \left( t-t^{\prime }\right) ,  \label{3-28}
\end{equation}%
\begin{equation}
\left\langle \pi _{i}(\overrightarrow{r},t)\pi _{j}(\overrightarrow{r}%
^{\prime },t^{\prime })\right\rangle =2k_{B}\left( T^{st}\right) ^{2}\kappa
_{ij}^{st}\delta \left( \overrightarrow{r}-\overrightarrow{r}^{\prime
}\right) \delta \left( t-t^{\prime }\right) ,  \label{3-29}
\end{equation}%
\begin{equation}
\left\langle \Upsilon _{\mu }(\overrightarrow{r},t)\Upsilon _{\nu }(%
\overrightarrow{r}^{\prime },t^{\prime })\right\rangle =2k_{B}T^{st}\frac{1}{%
\gamma _{1}}\left( \delta _{\mu \nu }^{\text{ }\perp }\right) ^{st}\delta
\left( \overrightarrow{r}-\overrightarrow{r}^{\prime }\right) \delta \left(
t-t^{\prime }\right) .  \label{3-30}
\end{equation}%
Here $k_{B}$\ is Boltzmann's constant and $T^{st}$ is given by Eq. (\ref{2-3}%
). Substitution of Eqs. (\ref{2-1})-(\ref{2-5}), (\ref{3-18}) and (\ref{3-20}%
) into Eqs. (\ref{3-14})-(\ref{3-17}), leads to the following first order in
the gradients set of linear, fluctuating nematodynamic equations 
\begin{equation}
\frac{\partial }{\partial t}\delta \rho =\rho _{0}\beta \left[ X+\frac{%
g\beta T_{0}}{\left( \gamma -1\right) c_{p}}\right] \delta v_{z}-\rho
_{0}\left\{ 1-\beta \left[ X+\frac{g\beta T_{0}}{\left( \gamma -1\right)
c_{p}}\right] z\right\} \nabla _{l}\delta v_{l},  \label{3-31}
\end{equation}%
\begin{gather}
\rho _{0}\left\{ 1-\beta \left[ X+\frac{g\beta T_{0}}{\left( \gamma
-1\right) c_{p}}\right] z\right\} \frac{\partial }{\partial t}\delta
v_{i}=-\nabla _{i}\delta p-\frac{1}{2}\lambda _{kji}^{st}\nabla _{j}\delta
h_{k}  \notag \\
+\nu _{ijlm}^{st}\nabla _{j}\partial _{m}\delta v_{l}-g\delta _{iz}\delta
\rho +\nabla _{j}\Sigma _{ij},  \label{3-32}
\end{gather}%
\begin{gather}
\rho _{0}T_{0}\left\{ 1-\beta \left[ X+\frac{g\beta T_{0}}{\left( \gamma
-1\right) c_{p}}\right] z\right\} \left( 1-\frac{\alpha }{T_{0}}z\right)
\left( \frac{\partial }{\partial t}\delta s+\frac{c_{p}}{T_{0}}X\delta
v_{z}\right)  \notag \\
=-\alpha \nabla _{l}\delta \kappa _{lz}+\kappa _{lj}^{st}\nabla _{l}\nabla
_{j}\delta T-\nabla _{l}\pi _{l},  \label{A33}
\end{gather}%
\begin{equation}
\frac{\partial }{\partial t}\delta n_{i}=\frac{1}{2}\lambda
_{ijk}^{st}\partial _{j}\delta v_{k}+\frac{1}{\gamma _{1}}\left( \delta
_{ik}^{\perp }\right) ^{st}\delta h_{k}+\Upsilon _{i}.  \label{3-34}
\end{equation}

A significant simplification of these equations is achived by noting that,
on the one hand, for a typical thermotropic nematic, $\rho _{0}\sim 1,$ $%
T_{0}\sim 10^{2},$ $\beta \sim 10^{-4},$ $c_{T}\sim 10^{5}$, $c_{p}\sim
10^{7}$ \cite{de gennes}. On the orher hand, in a typical light scattering
experiment $\alpha \leq 1$, $z\leq 1$, with $g\sim 10^{3}$. As a consequence 
$\beta Xz\lesssim 10^{-4}$, $\alpha z/T_{0}\lesssim 10^{-2}$, $g\beta
^{2}T_{0}z/\left[ \left( \gamma -1\right) c_{p}\right] =gz/c_{T}^{2}\sim
10^{-7}$; accordingly, in Eqs. (\ref{3-31})-(\ref{3-34}) the terms $\beta
Xz\lesssim 10^{-4}$, $gz/c_{T}^{2}\sim 10^{-7}$ and $\alpha z/T_{0}\lesssim
10^{-2}$ can be neglected. Thus, the set of equations (\ref{3-31})-(\ref%
{3-34}) becomes the more compact set of nematodynamic fluctuating equations 
\begin{equation}
\frac{\partial }{\partial t}\delta \rho =\rho _{0}\beta \left[ X+\frac{%
g\beta T_{0}}{\left( \gamma -1\right) c_{p}}\right] \delta v_{z}-\rho
_{0}\nabla _{l}\delta v_{l},  \label{3-35}
\end{equation}%
\begin{equation}
\rho _{0}\frac{\partial }{\partial t}\delta v_{i}=-\nabla _{i}\delta p-\frac{%
1}{2}\lambda _{kji}^{st}\nabla _{j}\delta h_{k}+\nu _{ijlm}^{st}\nabla
_{j}\nabla _{m}\delta v_{l}-g\delta _{iz}\delta \rho +\nabla _{j}\Sigma
_{ij},  \label{3-36}
\end{equation}%
\begin{equation}
\rho _{0}T_{0}\frac{\partial }{\partial t}\delta s=-\rho _{0}c_{p}X\delta
v_{z}-\alpha \nabla _{l}\delta \kappa _{lz}+\kappa _{lj}^{st}\nabla
_{l}\nabla _{j}\delta T-\nabla _{l}\pi _{l},  \label{3-37}
\end{equation}%
\begin{equation}
\frac{\partial }{\partial t}\delta n_{i}=\frac{1}{2}\lambda
_{ijk}^{st}\nabla _{j}\delta v_{k}+\frac{1}{\gamma _{1}}\left( \delta
_{ik}^{\perp }\right) ^{st}\delta h_{k}+\Upsilon _{i}.  \label{3-38}
\end{equation}

It is important emphasize several relevant features of these equations.
First, if the nematic is incompressible, $\nabla _{j}\delta v_{j}=0$, the
density fluctuations do not vanish due to the presence of the external
gradients. Secondly, they are consistent with known results in the
literature, specifically, in the absence of a gravitational field ($g=0$),
if the isotropic limit of (\ref{3-35})-(\ref{3-38}) is taken by setting $%
\delta n_{i}=0$ and if the term $g\rho _{0}\beta ^{2}T_{0}\delta v_{z}/\left[
\left( \gamma -1\right) c_{p}\right] =\rho _{0}g\delta v_{z}/c_{T}^{2}$ in
Eq. (\ref{3-35}) is eliminated, Eqs. (\ref{3-35})-(\ref{3-38}) reduce to the
corresponding hydrodynamic (non stochastic) equations for a simple fluid,
Eqs. (2) in Ref. \cite{lekkerkerker-boon}.

\subsection{Pressure-entropy representation}

Since the gravitational field induces a constant pressure gradient $\nabla
_{z}p$ given by Eq. (\ref{2-2}), and since in the geometry of the proposed
model the director field initially has a preferential orientation $\widehat{n%
}_{0}$ along the $z$ axis, the hydrodynamic variables may be divided into
two independent sets which are transverse and longitudinal to $\widehat{n}%
_{0}$ and the wave vector $\overrightarrow{k}$, also defined in Figure 1.
The former set is $\left\{ v_{x}\left( \overrightarrow{r},t\right)
,n_{x}\left( \overrightarrow{r},t\right) \right\} ,$ while the latter one is 
$\left\{ p\left( \overrightarrow{r},t\right) ,v_{y}\left( \overrightarrow{r}%
,t\right) ,v_{z}\left( \overrightarrow{r},t\right) ,s\left( \overrightarrow{r%
},t\right) ,n_{y}\left( \overrightarrow{r},t\right) \right\} $. The
corresponding linearized fluctuating hydrodynamic equations written in terms
of these sets of variables, are easily obtained by using the thermodynamic
relations%
\begin{equation}
\delta \rho =\left( \frac{\partial \rho }{\partial p}\right) _{s}^{st}\delta
p+\left( \frac{\partial \rho }{\partial s}\right) _{p}^{st}\delta s,
\label{5.24a}
\end{equation}%
\begin{equation}
\delta T=\left( \frac{\partial T}{\partial p}\right) _{s}^{st}\delta
p+\left( \frac{\partial T}{\partial s}\right) _{p}^{st}\delta s,
\label{5.24b}
\end{equation}%
with $\left( \partial \rho /\partial p\right) _{s}^{st}$ $\equiv 1/c_{s}^{2}$%
, $\left( \partial \rho /\partial s\right) _{p}^{st}\equiv -\beta ^{st}\rho
^{st}T^{st}/c_{p}=-\beta \rho _{0}T_{0}/c_{p}$, $\left( \partial T/\partial
p\right) _{s}^{st}=\beta T_{0}/(\rho _{0}c_{p})$, $\left( \partial
T/\partial s\right) _{p}^{st}\equiv T^{st}/c_{p}=T_{0}/c_{p}$, being $\beta
^{st}\equiv -1/\rho ^{st}\left( \partial \rho /\partial T\right)
_{p}^{st}=-1/\rho _{0}\left( \partial \rho /\partial T\right) _{p}$, $\chi
_{i}^{st}\equiv \kappa _{i}/\rho ^{st}c_{p}=\kappa _{i}/\left( \rho
_{0}c_{p}\right) $, for $i=\perp $,$\parallel $, $\kappa _{a}=\kappa
_{\parallel }-\kappa _{\perp }$, the thermal diffusivity coefficient. In
writing relations (\ref{5.24a})-(\ref{5.24b}), it has been assumed that $%
\rho ^{st}\simeq \rho _{0}$, $T^{st}\simeq T_{0}$ and that the thermodynamic
quantities $\beta $, $c_{p}$, $c_{T}$, $c_{s}$, $\kappa _{i}$ in the steady
state, have the same values as in equilibrium. In this representation the
complete set of linearized, fluctuating, hydrodynamic equations for $\left\{
\delta p,\delta s,\delta v_{i},\delta n_{i}\right\} $ is given by 
\begin{gather}
\frac{\partial }{\partial t}\delta p=\rho _{0}g\delta v_{z}+(\gamma -1)\left[
\chi _{\perp }\left( \nabla _{x}^{2}+\nabla _{y}^{2}\right) +\chi
_{\parallel }\nabla _{z}^{2}\right] \delta p+\frac{\rho _{0}}{\beta }(\gamma
-1)\left[ \chi _{\perp }\left( \nabla _{x}^{2}+\nabla _{y}^{2}\right) \right.
\notag \\
\left. +\chi _{\parallel }\nabla _{z}^{2}\right] \delta s-\rho
_{0}c_{s}^{2}\nabla _{i}\delta v_{i}-\alpha \beta \rho _{0}\chi
_{a}c_{s}^{2}\left( \nabla _{x}\delta n_{x}+\nabla _{y}\delta n_{y}\right) -%
\frac{\gamma -1}{\beta T_{0}}\nabla _{j}\pi _{j},  \label{5.24c}
\end{gather}%
\begin{gather}
\rho _{0}\frac{\partial }{\partial t}\delta v_{x}=-\nabla _{x}\delta p+\left[
(\nu _{2}+\nu _{4})\nabla _{x}^{2}+\nu _{2}\nabla _{y}^{2}+\nu _{3}\nabla
_{z}^{2}\right] \delta v_{x}+\nu _{4}\nabla _{x}\nabla _{y}\delta v_{y} 
\notag \\
+(\nu _{3}+\nu _{5})\nabla _{z}\nabla _{x}\delta v_{z}-\frac{1}{2}\left(
\lambda +1\right) (K_{1}\nabla _{x}^{2}+K_{2}\nabla _{y}^{2}+K_{3}\nabla
_{z}^{2})\nabla _{z}\delta n_{x}  \notag \\
-\frac{1}{2}\left( \lambda +1\right) \left( K_{1}\right. \left.
-K_{2}\right) \nabla _{z}\nabla _{x}\nabla _{y}\delta n_{y}+\nabla
_{j}\Sigma _{xj},  \label{5.24d}
\end{gather}%
\begin{gather}
\rho _{0}\frac{\partial }{\partial t}\delta v_{y}=-\nabla _{y}\delta p+\nu
_{4}\nabla _{y}\nabla _{x}\delta v_{x}+\left[ \nu _{2}\nabla _{x}^{2}+(\nu
_{2}+\nu _{4})\nabla _{y}^{2}+\nu _{3}\nabla _{z}^{2}\right] \delta v_{y} 
\notag \\
+(\nu _{3}+\nu _{5})\nabla _{z}\nabla _{y}\delta v_{z}-\frac{1}{2}\left(
\lambda +1\right) (K_{1}-K_{2})\nabla _{z}\nabla _{x}\nabla _{y}\delta n_{x}
\notag \\
-\frac{1}{2}\left( \lambda +1\right) (K_{2}\nabla _{x}^{2}+K_{1}\nabla
_{y}^{2}\left. +K_{3}\nabla _{z}^{2}\right) \nabla _{z}\delta n_{y}+\nabla
_{j}\Sigma _{yj},  \label{5.24dd}
\end{gather}%
\begin{gather}
\rho _{0}\frac{\partial }{\partial t}\delta v_{z}=-\nabla _{z}\delta p-\frac{%
g}{c_{s}^{2}}\delta p+(\nu _{3}+\nu _{5})\nabla _{z}\nabla _{x}\delta
v_{x}+(\nu _{3}+\nu _{5})\nabla _{z}\nabla _{y}\delta v_{y}+\left[ \nu
_{3}\left( \nabla _{x}^{2}+\nabla _{y}^{2}\right) \right.  \notag \\
+\left( 2\nu _{1}+\nu _{2}-\nu _{4}\right. \left. \left. +2\nu _{5}\right)
\nabla _{z}^{2}\right] \delta v_{z}-\frac{1}{2}\left( \lambda -1\right) %
\left[ K_{1}\left( \nabla _{x}^{2}+\nabla _{y}^{2}\right) +K_{3}\nabla
_{z}^{2}\right] \nabla _{x}\delta n_{x}  \notag \\
-\frac{1}{2}\left( \lambda -1\right) \left[ K_{1}\left( \nabla
_{x}^{2}+\nabla _{y}^{2}\right) \right. \left. +K_{3}\nabla _{z}^{2}\right]
\nabla _{y}\delta n_{y}+g\frac{\beta \rho _{0}T_{0}}{c_{p}}\delta s+\nabla
_{j}\Sigma _{zj},  \label{5.24e}
\end{gather}%
\begin{gather}
\frac{\partial }{\partial t}\delta s=-\frac{c_{p}}{T_{0}}X\delta v_{z}+\frac{%
\beta }{\rho _{0}}\left[ \chi _{\perp }\left( \nabla _{x}^{2}+\nabla
_{y}^{2}\right) +\chi _{\parallel }\nabla _{z}^{2}\right] \delta p+\left[
\chi _{\perp }\left( \nabla _{x}^{2}+\nabla _{y}^{2}\right) \right.  \notag
\\
\left. +\chi _{\parallel }\nabla _{z}^{2}\right] \delta s-\alpha \frac{\chi
_{a}c_{p}}{T_{0}}\left( \nabla _{x}\delta n_{x}+\nabla _{y}\delta
n_{y}\right) -\frac{1}{\rho _{0}T_{0}}\nabla _{j}\pi _{j},  \label{5.24ee}
\end{gather}%
\begin{gather}
\frac{\partial }{\partial t}\delta n_{x}=\frac{1}{2}\left( \lambda -1\right)
\nabla _{x}\delta v_{z}+\frac{1}{2}\left( \lambda +1\right) \nabla
_{z}\delta v_{x}+\frac{1}{\gamma _{1}}(K_{1}-K_{2})\nabla _{x}\nabla
_{y}\delta n_{y}  \notag \\
+\frac{1}{\gamma _{1}}(K_{1}\nabla _{x}^{2}+K_{2}\nabla _{y}^{2}+K_{3}\nabla
_{z}^{2})\delta n_{x}+\Upsilon _{x},  \label{5.24ff}
\end{gather}%
\begin{gather}
\frac{\partial }{\partial t}\delta n_{y}=\frac{1}{2}\left( \lambda +1\right)
\nabla _{z}\delta v_{y}+\frac{1}{2}\left( \lambda -1\right) \nabla
_{y}\delta v_{z}+\frac{1}{\gamma _{1}}(K_{1}-K_{2})\nabla _{x}\nabla
_{y}\delta n_{x}  \notag \\
+\frac{1}{\gamma _{1}}(K_{2}\nabla _{x}^{2}+K_{1}\nabla _{y}^{2}+K_{3}\nabla
_{z}^{2})\delta n_{y}+\Upsilon _{y},  \label{5.24f}
\end{gather}%
whith $j=x$, $y$, $z$, and where the $FDR$ of the stochastic components of
the fluxes are given by Eqs. (\ref{3-28})-(\ref{3-30}).

\subsection{Symmetry breaking representation}

For the purpose of calculating the spectrum of light scattering of the
nematic, it will be convenient to introduce a different set of fluctuating
thermodynamic variables that takes into account the effect of the intrinsec
anisotropy of the fluid. A proper set of new state variables that describe
the dynamics of fluctuations in a simple fluid with broken symmetry along
the $z$ axis, was proposed long ago in Ref. \cite{lekkerkerker-boon}.
Actually, this is the case of the nematic layer under consideration, because
owing to the initial orientation of the director $\widehat{n}_{i}^{st}$, the 
$NLC$ exhibits several symmetries, namely, rotational invariances around the 
$z$ axis, under inversions with respect to the $xy$ plane and at reflections
on planes containing the $z$ axis. Thus, following Ref. \cite%
{lekkerkerker-boon} we introduce the set of variables $\left\{ \delta
\varphi ,\delta \psi ,\delta \xi ,\delta f_{1},\delta f_{2},\delta s,\delta
p\right\} $ defined as follows \cite{Chandrasekhar (Hydro)},%
\begin{equation}
\delta \varphi \equiv \nabla \cdot \delta \overrightarrow{v}=\nabla
_{x}\delta v_{x}+\nabla _{y}\delta v_{y}+\nabla _{z}\delta v_{z},
\label{6.1}
\end{equation}%
\begin{equation}
\delta \psi \equiv \left( \nabla \times \delta \overrightarrow{v}\right)
_{z}=\nabla _{x}\delta v_{y}-\nabla _{y}\delta v_{x},  \label{6.2}
\end{equation}%
\begin{equation}
\delta \xi \equiv \left( \nabla \times \nabla \times \delta \overrightarrow{v%
}\right) _{z}=\frac{\partial \delta \varphi }{\partial z}-\left( \frac{%
\partial ^{2}}{\partial x^{2}}+\frac{\partial ^{2}}{\partial y^{2}}+\frac{%
\partial ^{2}}{\partial z^{2}}\right) \delta v_{z}.  \label{6.3}
\end{equation}%
By analogy, the director deviations $\delta \overrightarrow{n}$ are 
\begin{equation}
\delta f_{1}\equiv \nabla \cdot \delta \overrightarrow{n}=\nabla _{x}\delta
n_{x}+\nabla _{y}\delta n_{y},  \label{6.4}
\end{equation}%
\begin{equation}
\delta f_{2}\equiv \left( \nabla \times \delta \overrightarrow{n}\right)
_{z}=\nabla _{x}\delta n_{y}-\nabla _{y}\delta n_{x}.  \label{6.5}
\end{equation}

In this new representation the complete set of hydrodynamic equations (\ref%
{5.24c})-(\ref{5.24f}) takes the form

\begin{gather}
\frac{\partial }{\partial t}\delta p=(\gamma -1)(\chi _{\perp }\nabla
_{\perp }^{2}+\chi _{\parallel }\nabla _{\parallel }^{2})\delta p+\frac{\rho
_{0}}{\beta }(\gamma -1)(\chi _{\perp }\nabla _{\perp }^{2}+\chi _{\parallel
}\nabla _{\parallel }^{2})\delta s  \notag \\
+g\rho _{0}\delta v_{z}-\rho _{0}c_{s}^{2}\delta \varphi -\alpha \beta \rho
_{0}\chi _{a}c_{s}^{2}\delta f_{1}-\frac{\gamma -1}{\beta T_{0}}\nabla
_{j}\pi _{j},  \label{6.6}
\end{gather}%
\begin{gather}
\rho _{0}\frac{\partial }{\partial t}\delta \varphi =[(2\nu _{3}-\nu
_{2}-\nu _{4}+\nu _{5})\nabla _{\perp }^{2}+(2\nu _{1}+\nu _{2}-2\nu
_{3}-\nu _{4}+\nu _{5})\nabla _{\parallel }^{2}]\nabla _{z}\delta v_{z} 
\notag \\
-\left( \nabla ^{2}+\frac{g}{c_{s}^{2}}\nabla _{z}\right) \delta p+g\frac{%
\beta \rho _{0}T_{0}}{c_{p}}\nabla _{z}\delta s+\left[ (\nu _{2}+\nu
_{4})\nabla _{\perp }^{2}+(2\nu _{3}+\nu _{5})\nabla _{\parallel }^{2}\right]
\delta \varphi  \notag \\
-\lambda (K_{1}\nabla _{\perp }^{2}+K_{3}\nabla _{\parallel }^{2})\nabla
_{z}\delta f_{1}+\nabla _{j}\left( \nabla _{x}\Sigma _{xj}+\nabla _{y}\Sigma
_{yj}+\nabla _{z}\Sigma _{zj}\right) ,  \label{6.7}
\end{gather}%
\begin{gather}
\frac{\partial }{\partial t}\delta s=\frac{\beta }{\rho _{0}}\left( \chi
_{\perp }\nabla _{\perp }^{2}+\chi _{\parallel }\nabla _{\parallel
}^{2}\right) \delta p+\left( \chi _{\perp }\nabla _{\perp }^{2}+\chi
_{\parallel }\nabla _{\parallel }^{2}\right) \delta s  \notag \\
-\frac{c_{p}}{T_{0}}X\delta v_{z}-\alpha \frac{\chi _{a}c_{p}}{T_{0}}\delta
f_{1}-\frac{1}{\rho _{0}T_{0}}\nabla _{j}\pi _{j},  \label{6.8}
\end{gather}%
\begin{gather}
\rho _{0}\frac{\partial }{\partial t}\delta \xi =\frac{g}{c_{s}^{2}}\nabla
_{\perp }^{2}\delta p-g\frac{\beta \rho _{0}T_{0}}{c_{p}}\nabla _{\perp
}^{2}\delta s+\left[ (\nu _{2}-\nu _{3}+\nu _{4}-\nu _{5})\nabla _{\perp
}^{2}+\nu _{3}\nabla _{\parallel }^{2}\right] \nabla _{z}\delta \varphi 
\notag \\
+\frac{1}{2}[(\lambda -1)\nabla _{\perp }^{2}-(\lambda +1)\nabla _{\parallel
}^{2}](K_{1}\nabla _{\perp }^{2}+K_{3}\nabla _{\parallel }^{2})\delta
f_{1}-[\nu _{3}(\nabla _{\perp }^{2}-\nabla _{\parallel }^{2})^{2}  \notag \\
+2(\nu _{1}+\nu _{2})\nabla _{\perp }^{2}\nabla _{\parallel }^{2}]\delta
v_{z}+\nabla _{z}\nabla _{j}\left( \nabla _{x}\Sigma _{xj}+\nabla _{y}\Sigma
_{yj}+\nabla _{z}\Sigma _{zj}\right) -\nabla ^{2}\nabla _{j}\Sigma _{zj},
\label{6.9}
\end{gather}%
\begin{gather}
\frac{\partial }{\partial t}\delta f_{1}=\frac{1}{2}\left[ \left( \lambda
-1\right) \nabla _{\perp }^{2}-\left( \lambda +1\right) \nabla _{z}^{2}%
\right] \delta v_{z}+\frac{1}{\gamma _{1}}(K_{1}\nabla _{\perp
}^{2}+K_{3}\nabla _{\parallel }^{2})\delta f_{1}  \notag \\
+\frac{1}{2}\left( \lambda +1\right) \nabla _{z}\delta \varphi +\nabla
_{x}\Upsilon _{x}+\nabla _{y}\Upsilon _{y},  \label{6.10}
\end{gather}%
\begin{gather}
\rho _{0}\frac{\partial }{\partial t}\delta \psi =(\nu _{2}\nabla _{\perp
}^{2}+\nu _{3}\nabla _{\parallel }^{2})\delta \psi -\frac{1}{2}\left(
\lambda +1\right) (K_{2}\nabla _{\perp }^{2}+K_{3}\nabla _{\parallel
}^{2})\nabla _{z}\delta f_{2}  \notag \\
+\nabla _{j}\left( \nabla _{x}\Sigma _{yj}-\nabla _{y}\Sigma _{xj}\right) ,
\label{6.11}
\end{gather}%
\begin{gather}
\frac{\partial }{\partial t}\delta f_{2}=\frac{1}{2}\left( \lambda +1\right)
\nabla _{z}\delta \psi +\frac{1}{\gamma _{1}}(K_{2}\nabla _{\perp
}^{2}+K_{3}\nabla _{\parallel }^{2})\delta f_{2}  \notag \\
+\nabla _{x}\Upsilon _{y}-\nabla _{y}\Upsilon _{x},  \label{6.12}
\end{gather}%
where $\nabla _{\perp }^{2}\equiv \nabla _{x}^{2}+\nabla _{y}^{2},$ $\nabla
_{\parallel }^{2}\equiv \nabla _{z}^{2}$. In Eqs. (\ref{6.6})-(\ref{6.12}) $%
\delta v_{z}$ is coupled to $\delta \xi $ and $\delta \varphi $ through the
relation%
\begin{equation}
\delta \xi \equiv \nabla _{z}\delta \varphi -\nabla ^{2}\delta v_{z}.
\label{6.13}
\end{equation}

Furthermore, if the Fourier transform of an arbitrary field $A(%
\overrightarrow{r},t)$ with respect to $\overrightarrow{r}$ is defined by%
\begin{equation}
\widetilde{A}(\overrightarrow{k},\omega )\equiv \frac{1}{\left( 2\pi \right)
^{4}}\int_{-\infty }^{\infty }A(\overrightarrow{r},t)\exp \left[ -i(%
\overrightarrow{k}\cdot \overrightarrow{r}-\omega t)\right] d\overrightarrow{%
r}dt,  \label{5.28a}
\end{equation}%
with%
\begin{equation}
A(\overrightarrow{r},t)=\int_{-\infty }^{\infty }\widetilde{A}(%
\overrightarrow{k},\omega )\exp \left[ i(\overrightarrow{k}\cdot 
\overrightarrow{r}-\omega t)\right] d\overrightarrow{k}d\omega ,
\label{5.28b}
\end{equation}%
in matrix form the transformed set of Eqs. (\ref{6.6})-(\ref{6.12}) reads 
\begin{equation}
\frac{\partial }{\partial t}\delta \overrightarrow{X}(\overrightarrow{k}%
,t)=-M\delta \overrightarrow{X}(\overrightarrow{k},t)+\overrightarrow{\Theta 
}(\overrightarrow{k},t),  \label{5.29a}
\end{equation}%
where 
\begin{equation}
\delta \overrightarrow{X}(\overrightarrow{k},t)=\left( \delta 
\overrightarrow{X}^{L},\delta \overrightarrow{X}^{T}\right) ^{t}
\label{5.29aa}
\end{equation}%
and%
\begin{equation}
\delta \overrightarrow{X}^{L}(\overrightarrow{k},t)=\left( \delta \widetilde{%
p},\delta \widetilde{\varphi },\delta \widetilde{s},\delta \widetilde{\xi }%
,\delta \widetilde{f_{1}}\right) ^{t},  \label{5.29ab}
\end{equation}%
\begin{equation}
\delta \overrightarrow{X}^{T}(\overrightarrow{k},t)=\left( \delta \widetilde{%
\psi },\delta \widetilde{f_{2}}\right) ^{t}.  \label{5.29ac}
\end{equation}%
The hydrodynamic matrix $M$ is diagonal by blocks, 
\begin{equation}
M=\left( 
\begin{array}{c|c}
M^{L} & 0 \\ \hline
0 & M^{T}%
\end{array}%
\right) .  \label{5.29b}
\end{equation}%
where the superscripts $L$ and $T$\ denote, respectively, the longitudinal
and transverse sets of variables. The explicit form of the submatrices $%
M^{L} $, $M^{T}$, is 
\begin{equation}
M^{L}=\left( 
\begin{array}{ccccc}
\left( \gamma -1\right) D_{T}k^{2} & \rho _{0}c_{s}^{2}+g\frac{\rho
_{0}ik_{z}}{k^{2}} & \frac{\rho _{0}}{\beta }\left( \gamma -1\right)
D_{T}k^{2} & -g\frac{\rho _{0}}{k^{2}} & \alpha \beta \rho _{0}\chi
_{a}c_{s}^{2} \\ 
-\frac{k^{2}}{\rho _{0}}+g\frac{ik_{z}}{\rho _{0}c_{s}^{2}} & \sigma
_{1}k^{2} & -g\frac{\beta T_{0}}{c_{p}}ik_{z} & \sigma _{2}ik_{z} & -\frac{%
\lambda K_{I}}{\rho _{0}}ik^{2}k_{z} \\ 
\frac{\beta }{\rho _{0}}D_{T}k^{2} & -X\frac{c_{p}ik_{z}}{T_{0}k^{2}} & 
D_{T}k^{2} & X\frac{c_{p}}{T_{0}k^{2}} & \alpha \frac{\chi _{a}c_{p}}{T_{0}}
\\ 
g\frac{k_{\perp }^{2}}{\rho _{0}c_{s}^{2}} & -\sigma _{2}ik_{\perp }^{2}k_{z}
& -g\frac{\beta T_{0}}{c_{p}}k_{\perp }^{2} & \sigma _{3}k^{2} & -\frac{%
\Omega }{\rho _{0}}K_{I}k^{4} \\ 
0 & -\lambda \frac{ik_{\perp }^{2}k_{z}}{k^{2}} & 0 & \Omega & \frac{K_{I}}{%
\gamma _{1}}k^{2}%
\end{array}%
\right)  \label{5.29c}
\end{equation}%
and%
\begin{equation*}
M^{T}=\left( 
\begin{array}{cc}
\sigma _{4}k^{2} & -\frac{\lambda _{+}K_{II}}{\rho _{0}}ik^{2}k_{z} \\ 
-\lambda _{+}ik_{z} & \frac{K_{II}}{\gamma _{1}}k^{2}%
\end{array}%
\right) .
\end{equation*}%
with%
\begin{equation}
D_{T}\equiv \frac{1}{k^{2}}\left( \chi _{\perp }k_{\perp }^{2}+\chi
_{\parallel }k_{\parallel }^{2}\right) ,  \label{5.30a}
\end{equation}%
\begin{equation}
\sigma _{1}\equiv \frac{1}{\rho _{0}k^{4}}[\left( \nu _{2}+\nu _{4}\right)
k_{\perp }^{4}+2\left( 2\nu _{3}+\nu _{5}\right) k_{\parallel }^{2}k_{\perp
}^{2}+\left( 2\nu _{1}+\nu _{2}-\nu _{4}+2\nu _{5}\right) k_{\parallel
}^{4}],  \label{5.30b}
\end{equation}%
\begin{equation}
\sigma _{2}\equiv \frac{1}{\rho _{0}k^{2}}\left[ \left( -\nu _{2}+2\nu
_{3}-\nu _{4}+\nu _{5}\right) k_{\perp }^{2}+\left( 2\nu _{1}+\nu _{2}-2\nu
_{3}-\nu _{4}+\nu _{5}\right) k_{\parallel }^{2}\right] ,  \label{5.30c}
\end{equation}%
\begin{equation}
\sigma _{3}\equiv \frac{1}{\rho _{0}k^{4}}\left[ 2\left( \nu _{1}+\nu
_{2}\right) k_{\parallel }^{2}k_{\perp }^{2}+\nu _{3}\left( k_{\parallel
}^{2}-k_{\perp }^{2}\right) ^{2}\right] ,  \label{5.30d}
\end{equation}%
\begin{equation}
\sigma _{4}\equiv \frac{1}{\rho _{0}k^{2}}\left( \nu _{2}k_{\perp }^{2}+\nu
_{3}k_{\parallel }^{2}\right) ,  \label{5.30e}
\end{equation}%
\begin{equation}
K_{I}\equiv \frac{1}{k^{2}}\left( K_{1}k_{\perp }^{2}+K_{3}k_{\parallel
}^{2}\right) ,  \label{5.30f}
\end{equation}%
\begin{equation}
\text{\ }K_{II}\equiv \frac{1}{k^{2}}\left( K_{2}k_{\perp
}^{2}+K_{3}k_{\parallel }^{2}\right) ,  \label{5.30g}
\end{equation}%
\begin{equation}
\Omega \equiv \frac{1}{k^{2}}\left( \lambda _{-}k_{\perp }^{2}-\lambda
_{+}k_{\parallel }^{2}\right) ,  \label{5.30h}
\end{equation}%
\begin{equation}
\lambda _{-}\equiv \frac{1}{2}\left( \lambda -1\right) ,\text{ \ \ \ \ }%
\lambda _{+}\equiv \frac{1}{2}\left( \lambda +1\right) .  \label{5.30i}
\end{equation}%
It should be noted that $D_{T}$ has the dimensions and values of the orders
of magnitude of the coefficients of thermal diffusivity $\chi _{\perp },\chi
_{\parallel }$. The quantities $\sigma _{1}$, $\sigma _{2}$, $\sigma _{3}$, $%
\sigma _{4}$, have values comparable to the coefficients $\nu _{j}/\rho _{0}$
(for $j=1\ldots 5$); while $K_{I}$, $K_{II}$, have similar values to those
of the elastic constants $K_{j}$ (for $j=1$, $2$, $3$). Finally, the
dimensionless quantity $\Omega $ is a function of the previously defined
dimensionless coefficient $\lambda $ and is a measure of the anisotropy of
the nematic.

The statistical terms in Eq. (\ref{5.29a}) are given by the column vector%
\begin{equation}
\overrightarrow{\Theta }(\overrightarrow{k},t)=\left( \overrightarrow{\Theta 
}^{L},\overrightarrow{\Theta }^{T}\right) ^{t},  \label{5.31}
\end{equation}%
where the superscript $t$ denotes the transpose. Also%
\begin{equation}
\overrightarrow{\Theta }^{L}(\overrightarrow{k},t)=\left( 
\begin{array}{c}
-\frac{i\left( \gamma -1\right) }{\beta T_{0}}k_{j}\widetilde{\pi }_{j} \\ 
-\frac{k_{j}}{\rho _{0}}\left( k_{x}\widetilde{\Sigma }_{xj}+k_{y}\widetilde{%
\Sigma }_{yj}+k_{z}\widetilde{\Sigma }_{zj}\right) \\ 
-\frac{ik_{l}}{\rho _{0}T_{0}}\widetilde{\pi }_{l} \\ 
-\frac{ik_{z}k_{j}}{\rho _{0}}\left( k_{x}\widetilde{\Sigma }_{xj}+k_{y}%
\widetilde{\Sigma }_{yj}+k_{z}\widetilde{\Sigma }_{zj}\right) +\frac{ik^{2}}{%
\rho _{0}}k_{j}\widetilde{\Sigma }_{zj} \\ 
ik_{x}\Upsilon _{x}+ik_{y}\Upsilon _{y}%
\end{array}%
\right) ,  \label{5.31a}
\end{equation}%
\begin{equation}
\overrightarrow{\Theta }^{T}(\overrightarrow{k},t)=\left( 
\begin{array}{c}
-\frac{k_{j}}{\rho _{0}}\left( k_{x}\widetilde{\Sigma }_{yj}-k_{y}\widetilde{%
\Sigma }_{xj}\right) \\ 
ik_{x}\widetilde{\Upsilon }_{y}-ik_{y}\widetilde{\Upsilon }_{x}%
\end{array}%
\right) .  \label{5.31b}
\end{equation}

As a result of this change of representation, the original system of Eqs. (%
\ref{6.6})-(\ref{6.12}) is simplified into two uncoupled systems of
equations, namely, five equations for the longitudinal variables $\delta 
\overrightarrow{X}^{L}$, Eq. (\ref{5.29ab}), and two equations for the
transverse variables $\delta \overrightarrow{X}^{T}$, Eq. (\ref{5.29ac}).

\subsubsection{Equilibrium}

If the nonequilibrium terms containing $\alpha $ and $g$ are neglected in
Eqs. (\ref{5.29a}), the resulting equations describe the equilibrium state.
In this case the hydrodynamic matrix is given by%
\begin{equation}
M_{E}=\left( 
\begin{array}{c|c}
M_{L}^{E} & 0 \\ \hline
0 & M_{T}^{E}%
\end{array}%
\right) ,  \label{5.32a}
\end{equation}%
with%
\begin{equation}
M_{E}^{L}=\left( 
\begin{array}{ccccc}
\left( \gamma -1\right) D_{T}k^{2} & \rho _{0}c_{s}^{2} & \frac{\rho _{0}}{%
\beta }\left( \gamma -1\right) D_{T}k^{2} & 0 & 0 \\ 
-\frac{k^{2}}{\rho _{0}} & \sigma _{1}k^{2} & 0 & \sigma _{2}ik_{z} & -\frac{%
\lambda K_{I}}{\rho _{0}}ik^{2}k_{z} \\ 
\frac{\beta }{\rho _{0}}D_{T}k^{2} & 0 & D_{T}k^{2} & 0 & 0 \\ 
0 & -\sigma _{2}ik_{\perp }^{2}k_{z} & 0 & \sigma _{3}k^{2} & -\frac{\Omega 
}{\rho _{0}}K_{I}k^{4} \\ 
0 & -\lambda \frac{ik_{\perp }^{2}k_{z}}{k^{2}} & 0 & \Omega & \frac{K_{I}}{%
\gamma _{1}}k^{2}%
\end{array}%
\right)  \label{5.32b}
\end{equation}%
and%
\begin{equation}
M_{E}^{T}=\left( 
\begin{array}{cc}
\sigma _{4}k^{2} & -\frac{\lambda _{+}K_{II}}{\rho }ik^{2}k_{z} \\ 
-\lambda _{+}ik_{z} & \frac{K_{II}}{\gamma _{1}}k^{2}%
\end{array}%
\right) .  \label{5.32c}
\end{equation}%
Note that in Eq. (\ref{5.32a}) still prevails the same structure by blocks
shown in Eq. (\ref{5.29b}), that is, in the equilibrium state longitudinal
and transverse variables are completely decoupled; furthermore, $M_{E}^{L}$
contains more null entries and is simpler than $M^{L}$. On the other hand, $%
M_{E}^{T}$ is identical to $M^{T}$. Thus, the nonequilibrium effects caused
by the presence of $\alpha $ and $g$, only affect the longitudinal variables.

\section{Hydrodynamic modes}

In order to facilitate the calculation of hydrodynamic modes, we define the
following variables of the same dimension, $\left[ \delta z_{j}\right]
=M^{1/2}L^{-1/2}t$ (for $j=1$,$\ldots $, $7$),%
\begin{equation}
\begin{array}{ll}
z_{1}(\overrightarrow{k},t)\equiv \left( \frac{1}{\rho _{0}c_{s}^{2}}\right)
^{1/2}\delta \widetilde{p}, & \text{ \ \ \ }z_{5}(\overrightarrow{k}%
,t)\equiv \left( \frac{\rho _{0}c_{s}^{2}}{k^{2}}\right) ^{1/2}\delta 
\widetilde{f_{1}}, \\ 
z_{2}(\overrightarrow{k},t)\equiv \left( \frac{\rho _{0}}{k^{2}}\right)
^{1/2}\delta \widetilde{\varphi }, & \text{ \ \ \ }z_{6}(\overrightarrow{k}%
,t)\equiv \left( \frac{\rho _{0}}{k^{2}}\right) ^{1/2}\delta \widetilde{\psi 
}, \\ 
z_{3}(\overrightarrow{k},t)\equiv \left( \frac{\rho _{0}T_{0}}{c_{p}}\right)
^{1/2}\delta \widetilde{s}, & \text{ \ \ \ }z_{7}(\overrightarrow{k}%
,t)\equiv \left( \frac{\rho _{0}c_{s}^{2}}{k^{2}}\right) ^{1/2}\delta 
\widetilde{f_{2}}. \\ 
z_{4}(\overrightarrow{k},t)=\left( \frac{\rho _{0}}{k^{4}}\right)
^{1/2}\delta \widetilde{\xi }, & 
\end{array}
\label{5.33a}
\end{equation}%
The system (\ref{5.29a}) expressed in terms of the variables (\ref{5.33a})
is rewritten as%
\begin{equation}
\frac{\partial }{\partial t}\overrightarrow{Z}(\overrightarrow{k},t)=-N%
\overrightarrow{Z}(\overrightarrow{k},t)+\overrightarrow{\Xi }(%
\overrightarrow{k},t),  \label{5.33b}
\end{equation}%
in which 
\begin{equation}
\overrightarrow{Z}(\overrightarrow{k},t)=\left( \overrightarrow{Z}^{L},%
\overrightarrow{Z}^{T}\right) ^{t}  \label{5.33ba}
\end{equation}%
is the vector of variables of the same size, formed by the longitudinal%
\begin{equation}
\overrightarrow{Z}^{L}(\overrightarrow{k},t)=\left(
z_{1},z_{2},z_{3},z_{4},z_{5}\right) ^{t}  \label{5.33bc}
\end{equation}%
and transverse%
\begin{equation}
\overrightarrow{Z}^{T}(\overrightarrow{k},t)=\left( z_{6},z_{7}\right) ^{t}
\label{5.33bd}
\end{equation}%
variables. The hydrodynamic matrix $N$ 
\begin{equation}
N=\left( 
\begin{array}{c|c}
N^{L} & 0 \\ \hline
0 & N^{T}%
\end{array}%
\right)  \label{5.33c}
\end{equation}%
is composed by the submatrices%
\begin{equation}
N^{L}=\left( 
\begin{array}{ccccc}
\left( \gamma -1\right) D_{T}k^{2} & c_{s}k+\frac{g}{c_{s}}\frac{ik_{z}}{k}
& \left( \gamma -1\right) ^{1/2}D_{T}k^{2} & -\frac{g}{c_{s}} & \alpha \beta
\chi _{a}k \\ 
-c_{s}k+\frac{g}{c_{s}}\frac{ik_{z}}{k} & \sigma _{1}k^{2} & -\left( \gamma
-1\right) ^{1/2}\frac{g}{c_{s}}\frac{ik_{z}}{k} & \sigma _{2}ikk_{z} & -%
\frac{\lambda K_{I}}{\rho _{0}c_{s}}ik^{2}k_{z} \\ 
\left( \gamma -1\right) ^{1/2}D_{T}k^{2} & -\frac{\beta Xc_{s}}{\left(
\gamma -1\right) ^{1/2}}\frac{ik_{z}}{k} & D_{T}k^{2} & \frac{\beta Xc_{s}}{%
\left( \gamma -1\right) ^{1/2}} & \frac{\alpha \beta \chi _{a}}{\left(
\gamma -1\right) ^{1/2}}k \\ 
\frac{g}{c_{s}}\frac{k_{\perp }^{2}}{k^{2}} & -\sigma _{2}\frac{ik_{\perp
}^{2}k_{z}}{k} & -\left( \gamma -1\right) ^{1/2}\frac{g}{c_{s}}\frac{%
k_{\perp }^{2}}{k^{2}} & \sigma _{3}k^{2} & -\frac{\Omega K_{I}}{\rho
_{0}c_{s}}k^{3} \\ 
0 & -\lambda c_{s}\frac{ik_{\perp }^{2}k_{z}}{k^{2}} & 0 & \Omega c_{s}k & 
\frac{K_{I}}{\gamma _{1}}k^{2}%
\end{array}%
\right)  \label{5.33d}
\end{equation}%
and%
\begin{equation}
N^{T}=\left( 
\begin{array}{cc}
\sigma _{4}k^{2} & -\frac{\lambda _{+}K_{II}}{\rho _{0}c_{s}}ik^{2}k_{z} \\ 
-\lambda _{+}c_{s}ik_{z} & \frac{K_{II}}{\gamma _{1}}k^{2}%
\end{array}%
\right) .  \label{5.33e}
\end{equation}%
It can also easily verified that the dimension of each input $N_{ij}$ of the
matrix $N$ is $\left[ N_{ij}\right] =$ $t^{-1}$.

Moreover, in Eq. (\ref{5.33b}) the stochastic vectors 
\begin{equation}
\overrightarrow{\Xi }(\overrightarrow{k},t)=\left( \overrightarrow{\Xi }^{L},%
\overrightarrow{\Xi }^{T}\right) ^{t},  \label{5.33f}
\end{equation}%
are composed of the longitudinal%
\begin{equation}
\overrightarrow{\Xi }^{L}(\overrightarrow{k},t)=\left( \zeta _{1},\zeta
_{2},\zeta _{3},\zeta _{4},\zeta _{5}\right) ^{t}  \label{5.33g}
\end{equation}%
and the transverse%
\begin{equation}
\overrightarrow{\Xi }^{T}(\overrightarrow{k},t)=\left( \zeta _{6},\zeta
_{7}\right) ^{t}  \label{5.33h}
\end{equation}%
noise vectors. The stochastic noise components $\zeta _{m}$, $m=1\ldots 6$,
indicated in each one of them, are given explicitly in Appendix A. By taking
into account the fluctuation-dissipation relations Eqs. (\ref{3-28})-(\ref%
{3-29}), the autocorrelations and cross-correlations functions of the
stochastic noises, Eqs. (\ref{5.33ha})-(\ref{5.33hg}), averaged over the
steady state are also given in Appendix A.

In order to find hydrodynamic modes of the linear system (\ref{5.33b}), it
is required to calculate its eigenvalues ($\lambda $), which are given by
the roots of the characteristic equation%
\begin{equation}
p(\lambda )=p^{L}(\lambda )p^{T}(\lambda )=0,  \label{5.34}
\end{equation}%
where $p^{L}(\lambda )$ and $p^{T}(\lambda )$ are the characteristic
polynomials of fifth and second order in $\lambda $ of the matrices $N^{L}$
and $N^{T}$, respectively. These roots are calculated below.

\subsection{Longitudinal modes}

To simplify the calculation of $p_{L}(\lambda )$ of the matrix $N^{L}$ we
first define%
\begin{equation}
\overrightarrow{Z}^{L}=\left( \overrightarrow{Z}_{X}^{L},\overrightarrow{Z}%
_{Y}^{L}\right) ^{t}  \label{5.34aa}
\end{equation}%
with 
\begin{equation}
\overrightarrow{Z}_{X}^{L}=\left( z_{1},z_{2}\right) ^{t}  \label{5.34ab}
\end{equation}%
and 
\begin{equation}
\overrightarrow{Z}_{Y}^{L}=\left( z_{3},z_{4},z_{5}\right) ^{t}.
\label{5.34ac}
\end{equation}%
Then, from Eq. (\ref{5.33b}) we obtain the system%
\begin{equation}
\frac{\partial }{\partial t}\overrightarrow{Z}^{L}(\overrightarrow{k}%
,t)=-N^{L}\overrightarrow{Z}^{L}(\overrightarrow{k},t)+\overrightarrow{\Xi }%
^{L}(\overrightarrow{k},t),  \label{5.34a}
\end{equation}%
where the matrix of coefficients is%
\begin{equation}
N^{L}=\left( 
\begin{array}{c|c}
N_{XX}^{L} & N_{XY}^{L} \\ \hline
N_{YX}^{L} & N_{YY}^{L}%
\end{array}%
\right) .  \label{5.34b}
\end{equation}%
The submatrices are defined as%
\begin{equation}
N_{XX}^{L}=\left( 
\begin{array}{cc}
\left( \gamma -1\right) D_{T}k^{2} & c_{s}k+\frac{g}{c_{s}}\frac{ik_{z}}{k}
\\ 
-c_{s}k+\frac{g}{c_{s}}\frac{ik_{z}}{k} & \sigma _{1}k^{2}%
\end{array}%
\right) ,  \label{5.34c}
\end{equation}%
\begin{equation}
N_{XY}^{L}=\left( 
\begin{array}{ccc}
\left( \gamma -1\right) ^{1/2}D_{T}k^{2} & -\frac{g}{c_{s}} & \alpha \beta
\chi _{a}k \\ 
-\left( \gamma -1\right) ^{1/2}\frac{g}{c_{s}}\frac{ik_{z}}{k} & \sigma
_{2}ikk_{z} & -\frac{\lambda K_{I}}{\rho _{0}c_{s}}ik^{2}k_{z}%
\end{array}%
\right) ,  \label{5.34d}
\end{equation}%
\begin{equation}
N_{YX}^{L}=\left( 
\begin{array}{cc}
\left( \gamma -1\right) ^{1/2}D_{T}k^{2} & -\frac{\beta Xc_{s}}{\left(
\gamma -1\right) ^{1/2}}\frac{ik_{z}}{k} \\ 
\frac{g}{c_{s}}\frac{k_{\perp }^{2}}{k^{2}} & -\sigma _{2}\frac{ik_{\perp
}^{2}k_{z}}{k} \\ 
0 & -\lambda c_{s}\frac{ik_{\perp }^{2}k_{z}}{k^{2}}%
\end{array}%
\right)  \label{5.34e}
\end{equation}%
and%
\begin{equation}
N_{YY}^{L}=\left( 
\begin{array}{ccc}
D_{T}k^{2} & \frac{\beta Xc_{s}}{\left( \gamma -1\right) ^{1/2}} & \frac{%
\alpha \beta \chi _{a}}{\left( \gamma -1\right) ^{1/2}}k \\ 
-\left( \gamma -1\right) ^{1/2}\frac{g}{c_{s}}\frac{k_{\perp }^{2}}{k^{2}} & 
\sigma _{3}k^{2} & -\frac{\Omega K_{I}}{\rho _{0}c_{s}}k^{3} \\ 
0 & \Omega c_{s}k & \frac{K_{I}}{\gamma _{1}}k^{2}%
\end{array}%
\right) .  \label{5.34f}
\end{equation}%
Furthermore,%
\begin{equation}
\overrightarrow{\Xi }^{L}(\overrightarrow{k},t)=\left( \overrightarrow{\Xi }%
_{X}^{L},\overrightarrow{\Xi }_{Y}^{L}\right) ^{t}  \label{5.34fa}
\end{equation}%
is the vector of longitudinal stochastic terms, with components%
\begin{equation}
\overrightarrow{\Xi }_{X}^{L}(\overrightarrow{k},t)=\left( \zeta _{1},\zeta
_{2}\right) ^{t}  \label{5.34g}
\end{equation}%
and%
\begin{equation}
\overrightarrow{\Xi }_{Y}^{L}(\overrightarrow{k},t)=\left( \zeta _{3},\zeta
_{4},\zeta _{5}\right) ^{t}.  \label{5.34h}
\end{equation}

Following the method proposed by \cite{law-sengers} for a simple fluid, it
can be shown the system Eq. (\ref{5.34a}) has the property that, within a
very good approximation, the variables $\delta \overrightarrow{Z}_{X}^{L}$
and $\delta \overrightarrow{Z}_{Y}^{L}$ are mutually independent \cite{tesis}%
. This statement implies that in the matrix $N^{L}$ the blocks $N_{XY}^{L}$
and $N_{YX}^{L}$ can be neglected and Eq. (\ref{5.34b}) is simplified to%
\begin{equation}
N^{L}=\left( 
\begin{array}{c|c}
N_{XX}^{L} & 0 \\ \hline
0 & N_{YY}^{L}%
\end{array}%
\right) .  \label{5.35a}
\end{equation}%
Consequently, the set of equations (\ref{5.34a}) is reduced to the uncoupled
system%
\begin{equation}
\frac{\partial }{\partial t}\overrightarrow{Z}_{X}^{L}(\overrightarrow{k}%
,t)=-N_{XX}^{L}\overrightarrow{Z}_{X}^{L}(\overrightarrow{k},t)+%
\overrightarrow{\Xi }_{X}^{L}(\overrightarrow{k},t),  \label{5.35b}
\end{equation}%
\begin{equation}
\frac{\partial }{\partial t}\overrightarrow{Z}_{Y}^{L}(\overrightarrow{k}%
,t)=-N_{YY}^{L}\overrightarrow{Z}_{Y}^{L}(\overrightarrow{k},t)+%
\overrightarrow{\Xi }_{Y}^{L}(\overrightarrow{k},t).  \label{5.35c}
\end{equation}%
The same approximatiom allows to rewrite the characteristic polynomial of
longitudinal variables as%
\begin{equation}
p^{L}(\lambda )=p_{XX}^{L}(\lambda )p_{YY}^{L}(\lambda ),  \label{5.36}
\end{equation}%
where 
\begin{gather}
p_{XX}^{L}(\lambda )=\lambda ^{2}-\left[ \left( \gamma -1\right)
D_{T}k^{2}+\sigma _{1}k^{2}\right] \lambda  \notag \\
+\left( \gamma -1\right) \sigma _{1}k^{2}D_{T}k^{2}+k^{2}c_{s}^{2}+\frac{%
g^{2}}{c_{s}^{2}}\frac{k_{z}^{2}}{k^{2}}  \label{5.37aa}
\end{gather}%
and%
\begin{gather}
p_{YY}^{L}(\lambda )=\lambda ^{3}-\left( D_{T}k^{2}+\sigma _{3}k^{2}+\frac{%
K_{I}k^{2}}{\gamma _{1}}\right) \lambda ^{2}+\left( D_{T}k^{2}\sigma
_{3}k^{2}+D_{T}k^{2}\frac{K_{I}k^{2}}{\gamma _{1}}\right.  \notag \\
\left. +\sigma _{3}k^{2}\frac{K_{I}k^{2}}{\gamma _{1}}+\frac{\Omega
^{2}K_{I}k^{4}}{\rho _{0}}+gX\beta \frac{k_{\perp }^{2}}{k^{2}}\right)
\lambda -D_{T}k^{2}\sigma _{3}k^{2}\frac{K_{I}k^{2}}{\gamma _{1}}  \notag \\
-D_{T}k^{2}\frac{\Omega ^{2}K_{I}k^{4}}{\rho _{0}}-gX\beta \frac{k_{\perp
}^{2}}{k^{2}}\frac{K_{I}k^{2}}{\gamma _{1}}+g\alpha \beta \frac{k_{\perp
}^{2}}{k^{2}}\Omega \chi _{a}k^{2}.  \label{5.37ab}
\end{gather}

While there is no analytical difficulty to solve the quadratic and cubic
equations (\ref{5.37aa}) and (\ref{5.37ab}), the explicit form of their
exact roots can be quite complicated, especially for the latter. However, it
is possible to estimate them following a procedure based partially on a
method suggested in Ref. \cite{Mountain}. According to it, in Eq. (\ref%
{5.37aa}) the following quantities $\left( \gamma -1\right) D_{T}k^{2}$, $%
\sigma _{1}k^{2}$, $k^{2}c_{s}^{2}$ y $g^{2}k_{z}^{2}/(c_{s}^{2}k^{2})$, may
be identified. They depend on the thermal diffusion coefficient $D_{T}$, the
viscosity $\sigma _{1}$, as well as on the gravitational field $g$ and the
adiabatic speed of sound propagation $c_{s}$. On the other hand, in Eq. (\ref%
{5.37ab}) the quantities $g\alpha \beta \frac{k_{\perp }^{2}}{k^{2}}$, $%
gX\beta \frac{k_{\perp }^{2}}{k^{2}}$, $D_{T}k^{2}$, $\Omega \chi _{a}k^{2}$%
, $\sigma _{3}k^{2}$, $\frac{K_{1}}{\gamma _{1}}k^{2}$ and $\frac{\Omega
^{2}K_{I}}{\rho _{0}}k^{4}$, may be also identified. They depend on both
the, nematic material parameters, as the coefficients of thermal diffusivity 
$\chi _{\Vert }$, $\chi _{\bot }$, the viscosity coefficient $\nu _{3}$, the
elastic constants $K_{1}$,$K_{3}$, as well as on the temperature gradient $%
\alpha $ and the gravitational field $g$. It is helpful to compare these
quantities with $\omega \equiv c_{s}k$, by introducing the small or reducted
quantities%
\begin{gather}
a_{0}\equiv \frac{g\alpha \beta }{\omega }\frac{k_{\perp }^{2}}{k^{2}},\text{
\ }a_{0}^{\prime }\equiv \frac{gX\beta }{\omega }\frac{k_{\perp }^{2}}{k^{2}}%
,\text{ \ \ }a_{0}^{\prime \prime }\equiv \frac{g^{2}k_{z}^{2}}{\omega
c_{s}^{2}k^{2}},\text{ \ }a_{1}\equiv \frac{D_{T}k^{2}}{\omega },\text{ \ \ }%
a_{1}^{\prime }\equiv \frac{\Omega \chi _{a}k^{2}}{\omega },  \notag \\
a_{2}\equiv \frac{\sigma _{1}k^{2}}{\omega }\,,\text{ \ }a_{3}\equiv \frac{%
\sigma _{3}k^{2}}{\omega },\text{ \ \ }a_{5}\equiv \frac{K_{I}k^{2}}{\gamma
_{1}\omega }\,,\text{ \ \ }a_{6}\equiv \frac{\Omega ^{2}K_{I}k^{4}}{\rho
_{0}\omega }.  \label{5.37ac}
\end{gather}%
For most nematics at ambient temperatures, $\rho _{0}$ and $\Omega $ are of
order of magnitude $1$, $\gamma _{1}\sim 10^{-1}$ , $\chi _{i}$ and $\nu
_{i} $ are of order $10^{-2}$ $-$ $10^{-3}$, $K_{i}$ $\sim 10^{-6}$ $-$ $%
10^{-7}$, while $\beta \sim 10^{-4}$ \cite{de gennes}; also, we consider
that $\alpha \lesssim 1$ and $g\sim 10^{3}$. Since in typical light
scattering experiments $k=10^{5}cm^{-1}$ and $c_{s}=1.5\times 10^{5}cms^{-1}$
\cite{Berne-Pecora}, \cite{Boon and Yip}, the quantities given in Eq. (\ref%
{5.37ac}) have the following orders of magnitude: $a_{0}\sim 10^{-11}$, $%
a_{0}^{\prime }\sim 10^{-11}$, $a_{0}^{\prime \prime }\sim 10^{-14}$, $%
a_{1}\sim 10^{-3}$, $a_{1}^{\prime }\sim 10^{-3}$, $a_{2}\sim 10^{-2}$, $%
a_{3}\sim 10^{-2}$, $a_{5}\sim 10^{-5}$ and\ $a_{6}\sim 10^{4}$. If we were
to follow the method of Ref. \cite{Mountain}, the solutions of Eqs. (\ref%
{5.37aa}) and (\ref{5.37ab}) should be obtained by a perturbative
approximation in terms of these small quantities. However, we will improve
this approximation by using the exact roots of Eqs. (\ref{5.37aa}) and (\ref%
{5.37ab}) and by expressing them in terms of reduced quantities (\ref{5.37ac}%
) of lower order in $k^{2}$ \cite{tesis}. This procedure will be implemented
in the next two subsections.

\subsubsection{Sound longitudinal modes}

In accordance with Eq. (\ref{5.37aa}), the sound propagation modes are the
roots of the characteristic equation $p_{XX}^{L}(\lambda )=0$. In terms of
the variable $s\equiv \lambda /\omega $ and the small quantities given in
Eq. (\ref{5.37ac}), this characteristic equation is rewritten as%
\begin{equation}
s^{2}+A^{\prime }s+B^{\prime }=0,  \label{5.40ac}
\end{equation}%
where%
\begin{equation}
A^{\prime }\equiv -\left[ \left( \gamma -1\right) a_{1}+a_{2}\right] ,
\label{5.40ad}
\end{equation}%
\begin{equation}
B^{\prime }\equiv 1+\left( \gamma -1\right) a_{1}a_{2}+\frac{a_{0}^{\prime
\prime }}{\omega }.  \label{5.40ae}
\end{equation}%
Analytical solutions of Eq. (\ref{5.40ac}) are%
\begin{equation}
s_{+}\simeq -\frac{1}{2}A^{\prime }+\frac{1}{2}\sqrt{\Delta ^{\prime }},
\label{5.40af}
\end{equation}%
\begin{equation}
s_{-}\simeq -\frac{1}{2}A^{\prime }-\frac{1}{2}\sqrt{\Delta ^{\prime }},
\label{5.40ag}
\end{equation}%
in which 
\begin{equation}
\Delta ^{\prime }\equiv A^{\prime 2}-4B^{\prime }  \label{5.40ah}
\end{equation}%
is the discriminant. Its sign determines the nature of the roots (\ref%
{5.40af}) and (\ref{5.40ag}), which can only present one of the following
three characteristics: two real and distinct roots, if $\Delta ^{\prime }>0$%
; two real and equal roots, if $\Delta ^{\prime }=0$ and two complex
conjugate roots, when $\Delta ^{\prime }<0$. Thus, according to the orders
of magnitude of small amounts (\ref{5.37ac}) in the coefficients (\ref%
{5.40ad}) and (\ref{5.40ae}), the discriminant (\ref{5.40ah}) can be
simplified to $\Delta ^{\prime }\simeq -4k^{2}c_{s}^{2}$, given that $a_{2}$%
, $a_{1}$, $a_{0}^{\prime \prime }/\omega ^{2}\ll 1$. In fact, $\Delta
^{\prime }<0$ always. Note that since $a_{0}^{\prime \prime }/\omega \sim
10^{-24}$, the effect of external gravitational field $g$ in $\Delta
^{\prime }$ is negligible. Therefore, solutions (\ref{5.40af}) and (\ref%
{5.40ag}) will be complex conjugate,%
\begin{equation}
s_{+}\simeq \frac{1}{2}\left[ \left( \gamma -1\right) a_{1}+a_{2}\right] +i,
\label{5.40ai}
\end{equation}%
\begin{equation}
s_{-}\simeq \frac{1}{2}\left[ \left( \gamma -1\right) a_{1}+a_{2}\right] -i.
\label{5.40aj}
\end{equation}%
Rewriting these roots in terms of the variables $\lambda _{i}$ by means of
the relation $\lambda \equiv \omega s$, leads to%
\begin{equation}
\lambda _{1}\simeq \Gamma k^{2}+ic_{s}k,  \label{5.40ak}
\end{equation}%
\begin{equation}
\lambda _{2}\simeq \Gamma k^{2}-ic_{s}k,  \label{5.40al}
\end{equation}%
where 
\begin{equation}
\Gamma \equiv \frac{1}{2}\left[ \left( \gamma -1\right) D_{T}+\sigma _{1}%
\right]  \label{5.40am}
\end{equation}%
is the sound attenuation coefficient of the nematic fluid. It should be
noted that the sound propagation modes found, Eqs. (\ref{5.40ak}) and (\ref%
{5.40al}), are in complete agreement with those already reported in the
literature for $NLC$ \cite{Rusos}, \cite{grupo de Orsey}.

\subsubsection{Thermal diffusive, shear and director longitudinal modes}

According to Eq. (\ref{5.36}), the thermal diffusive, shear and director
modes, are the roots of the characteristic equation $p_{YY}^{L}(\lambda )=0$%
. Again, in terms of the variable $s\equiv \lambda /\omega $ and the small
quantities (\ref{5.37ac}), this equation reads%
\begin{equation}
s^{3}+As^{2}+Bs+C=0,  \label{5.37c}
\end{equation}%
where%
\begin{equation}
A\equiv -a_{1}-a_{3}-a_{5},  \label{5.37d}
\end{equation}%
\begin{equation}
B\equiv a_{1}a_{3}+a_{1}a_{5}+a_{3}a_{5}+\frac{a_{6}}{\omega }+\frac{%
a_{0}^{\prime }}{\omega },  \label{5.37e}
\end{equation}%
\begin{equation}
C\equiv -a_{1}a_{3}a_{5}-\frac{a_{1}a_{6}}{\omega }-\frac{a_{0}^{\prime
}a_{5}}{\omega }+\frac{a_{0}a_{1}^{\prime }}{\omega }.  \label{5.37f}
\end{equation}%
It may be noted that all terms present in the coefficients of the cubic
equation (\ref{5.37c}), given by Eqs. (\ref{5.37d})-(\ref{5.37f}), are lower
than unity. The exact solutions of the cubic equation (\ref{5.37c}) are%
\begin{equation}
s_{1}=-\frac{A}{3}-\frac{\sqrt[3]{2}\left( 3B-A^{2}\right) }{3F}+\frac{F}{3%
\sqrt[3]{2}},  \label{5.37g}
\end{equation}%
\begin{equation}
s_{2}=-\frac{A}{3}+\frac{\left( 1+i\sqrt{3}\right) \left( 3B-A^{2}\right) }{%
2^{2/3}3F}-\frac{\left( 1-i\sqrt{3}\right) F}{6\sqrt[3]{2}},  \label{5.37h}
\end{equation}%
\begin{equation}
s_{3}=-\frac{A}{3}+\frac{\left( 1-i\sqrt{3}\right) \left( 3B-A^{2}\right) }{%
2^{2/3}3F}-\frac{\left( 1+i\sqrt{3}\right) F}{6\sqrt[3]{2}},  \label{5.37i}
\end{equation}%
where%
\begin{equation}
F\equiv \sqrt[3]{-2A^{3}+9AB-27C+3\sqrt{3}\sqrt{\Delta }}  \label{5.37j}
\end{equation}%
with the discriminant 
\begin{equation}
\Delta \equiv -A^{2}B^{2}+4B^{3}+4A^{3}C-18ABC+27C^{2}.  \label{5.37k}
\end{equation}%
The sign of $\Delta $ determines the nature of the roots (\ref{5.37g})-(\ref%
{5.37i}); only one of the following three cases is possible: one real and
two complex conjugate roots, if $\Delta <0$; three real and distinct roots,
if $\Delta >0$ and three real roots, one different and two identical, if $%
\Delta =0$. Taking into account the orders of magnitude of small quantities (%
\ref{5.37ac}), the explicit expressions of the three roots Eqs. (\ref{5.37g}%
)-(\ref{5.37i}), are given up to first order in the small quantities, i. e.,
up to $k^{2}$ order, as%
\begin{equation}
s_{3,4}\simeq \frac{1}{2}\left( a_{1}+a_{3}-\frac{a_{6}}{\omega a_{3}}%
\right) \mp \frac{1}{2}\sqrt{\left( a_{1}+a_{3}-\frac{a_{6}}{\omega a_{3}}%
\allowbreak \right) ^{2}-4a_{1}a_{3}\left( 1-\frac{R}{R_{c}}\right) },
\label{5.38a}
\end{equation}%
\begin{equation}
s_{5}\simeq a_{5}+\frac{a_{6}}{\omega a_{3}},  \label{5.38b}
\end{equation}%
\begin{equation}
\frac{R}{R_{c}}\equiv -\left( \frac{a_{0}^{\prime }}{\omega a_{1}a_{3}}+%
\frac{a_{0}a_{1}^{\prime }}{\omega a_{1}^{2}a_{3}}+\frac{a_{0}a_{1}^{\prime }%
}{\omega a_{1}a_{3}^{2}}\right) .  \label{5.38d}
\end{equation}%
In Eq. (\ref{5.38d}) $R$ is the Rayleigh number and $R_{c}$ denotes its
critical value. The radicand of (\ref{5.38a}) is the discriminant 
\begin{equation}
\Delta \equiv \left( a_{1}+a_{3}-\frac{a_{6}}{\omega a_{3}}\allowbreak
\right) ^{2}-4a_{1}a_{3}\left( 1-\frac{R}{R_{c}}\right) .  \label{5.38e}
\end{equation}%
It should be noted that according to the orders of magnitude of the
quantities (\ref{5.37ac}) contained in (\ref{5.38a}) and (\ref{5.38b}), from
Eq. (\ref{5.38e}) it follows that $\Delta >0$. Consequently, the roots (\ref%
{5.38a})-(\ref{5.38b}) are real and distinct. Up to first order in the
amounts (\ref{5.37ac}), these roots are rewritten in terms of the variables $%
\lambda _{i}$ as%
\begin{align}
\lambda _{3,4}& \simeq \frac{1}{2}\left( D_{T}k^{2}+\sigma _{3}k^{2}-\frac{%
\Omega ^{2}K_{I}k^{4}}{\rho _{0}\sigma _{3}k^{2}}\right)  \notag \\
& \mp \frac{1}{2}\sqrt{\left( D_{T}k^{2}+\sigma _{3}k^{2}-\frac{\Omega
^{2}K_{I}k^{4}}{\rho _{0}\sigma _{3}k^{2}}\allowbreak \right)
^{2}-4D_{T}k^{2}\sigma _{3}k^{2}\left( 1-\frac{R}{R_{c}}\right) },
\label{5.38h}
\end{align}%
\begin{equation}
\lambda _{5}\simeq \frac{K_{I}k^{2}}{\gamma _{1}}+\frac{\Omega ^{2}K_{I}k^{4}%
}{\rho _{0}\sigma _{3}k^{2}},  \label{5.38i}
\end{equation}%
where Eq. (\ref{5.38d}) have been rewritten in the form%
\begin{equation}
\frac{R}{R_{c}}\equiv -\left[ \frac{gX\beta \frac{k_{\perp }^{2}}{k^{2}}}{%
D_{T}k^{2}\sigma _{3}k^{2}}+\frac{g\alpha \beta \frac{k_{\perp }^{2}}{k^{2}}%
\Omega \chi _{a}k^{2}}{\left( D_{T}k^{2}\right) ^{2}\sigma _{3}k^{2}}+\frac{%
g\alpha \beta \frac{k_{\perp }^{2}}{k^{2}}\Omega \chi _{a}k^{2}}{%
D_{T}k^{2}\left( \sigma _{3}k^{2}\right) ^{2}}\right] .  \label{5.38l}
\end{equation}%
Equation (\ref{5.38h}) corresponds to a pair of visco-heat modes which
result from the coupling between the thermal and shear modes. Their
existence is entirely due to the presence of both, the uniform temperature
gradient and the constant gravitational field, or only the gravity field.
According to the orders of magnitude of the material proerties and
experimental parameters indicated in Eq. (\ref{5.38l}), the first term is of
order $10^{-15}$, whereas the second and third terms are of $10^{-16}$.
Also, the discriminant Eq. (\ref{5.38e}) takes the form%
\begin{equation}
\Delta \equiv \left( D_{T}k^{2}+\sigma _{3}k^{2}-\frac{\Omega ^{2}K_{I}k^{4}%
}{\rho _{0}\sigma _{3}k^{2}}\right) ^{2}-4D_{T}k^{2}\sigma _{3}k^{2}\left( 1-%
\frac{R}{R_{c}}\right) .  \label{5.38k}
\end{equation}%
In Eq. (\ref{5.38l}) the presence of $\chi _{a}$ in the second and third
terms is indicative that the system under study is a nematic; besides, such
term is an order of magnitude greater than the first. If in this same
expression $\chi _{a}=0$, which occurs in the isotropic limit, then 
\begin{equation}
\frac{R}{R_{c}}\equiv -\frac{gX\beta }{D_{T}k^{2}\sigma _{3}k^{2}}\frac{%
k_{\perp }^{2}}{k^{2}},  \label{5.38la}
\end{equation}%
which has the same structure of the corresponding expression reported for a
simple fluid \cite{ortiz}, \cite{segre}, \cite{lekkerkerker-boon}. This
results allow us to quantify the effect produced in the modes, by $\alpha $
and $g$. Their influence is due to the coupling of the small quantities
present in the factor $g\alpha \beta \frac{k_{\perp }^{2}}{k^{2}}$. Also, it
is worth noting that the presence of $\alpha $ and $g$ produces a coupling
between the thermal and shear difusive modes, as may well be seen in Eq. (%
\ref{5.38h}).

\subsection{Values of $R$}

\subsubsection{Critical value\emph{\ }$(R=R_{c})$}

Some special values of $R$\ are of particular interest. For instance, if $R$
reaches its critical value $R_{c}$, then $\Delta =\left( D_{T}k^{2}+\sigma
_{3}k^{2}-\frac{\Omega ^{2}K_{I}k^{4}}{\rho _{0}\sigma _{3}k^{2}}\right)
^{2} $ and hence the modes (\ref{5.38h}) and (\ref{5.38i}) are simplified in
the form%
\begin{equation}
\lambda _{3}\simeq 0,  \label{5.38ra}
\end{equation}%
\begin{equation}
\lambda _{4}\simeq D_{T}k^{2}+\sigma _{3}k^{2}-\frac{\Omega ^{2}K_{I}k^{4}}{%
\rho _{0}\sigma _{3}k^{2}},  \label{5.38rb}
\end{equation}%
\begin{equation}
\lambda _{5}\simeq \frac{K_{I}k^{2}}{\gamma _{1}}+\frac{\Omega ^{2}K_{I}k^{4}%
}{\rho _{0}\sigma _{3}k^{2}},  \label{5.38rc}
\end{equation}%
which are in agreement with those reported in literature in this limit \cite%
{miyakawa}, \cite{migranov}. In this situation $\lambda _{3}$ vanishes, $%
\lambda _{5}$ is virtually unchanged, while $\lambda _{4}$ has contributions
from the thermal and shear difusive modes. It should be pointed out that
this phenomenon also occurs in the simple fluid, where there are two
diffusive modes, one of them also vanishes, and the other one has
contributions from the shear and thermal modes \cite{ortiz}, \cite%
{lekkerkerker-boon}.\ For a simple fluid, these features have been predicted
theoretically, and even more, corroborated experimentally. These results
suggest that it might be feasible to verify them experimentally also for
nematics. It should be stressed that the results obtained in this limit do
not coincide with those reported for a $NLC$, according to which the
director mode tends to zero, the shear mode does not change and there is an
additional mode which is the sum of the thermal and director modes \cite%
{miyakawa}, \cite{migranov}.

\subsubsection{Equilibrium state ($R=0)$}

In the absence of temperature the gradient $\alpha $ and the gravitational
field $g$, $R=0$ and from Eqs. (\ref{5.38h}) and (\ref{5.38i}) the
corresponding expressions for the thermal, shear and director diffusive
modes in the equilibrium state (identified by the superscript $e$) are
readily obtained. The correponding expressions reduce to%
\begin{equation}
\lambda _{3}^{e}\simeq D_{T}k^{2},  \label{5.38sa}
\end{equation}%
\begin{equation}
\lambda _{4}^{e}\simeq \sigma _{3}k^{2}-\frac{\Omega ^{2}K_{I}k^{4}}{\rho
_{0}\sigma _{3}k^{2}},  \label{5.38sb}
\end{equation}%
\begin{equation}
\lambda _{5}^{e}\simeq \frac{K_{I}k^{2}}{\gamma _{1}}+\frac{\Omega
^{2}K_{I}k^{4}}{\rho _{0}\sigma _{3}k^{2}},  \label{5.38sc}
\end{equation}%
which are well known results in the literature \cite{buka}, \cite{Rusos}, 
\cite{Leslie}. In this homogeneous thermodynamic equilibrium state, the
decay rates Eqs. (\ref{5.38sa})-(\ref{5.38sc}) are purely diffusive.

\subsubsection{Visco-heat propagation modes}

It has already been mentioned that owing to the orders of magnitude of the
small quantities (\ref{5.37ac}), the roots (\ref{5.38a}) and (\ref{5.38b})
are real and different. Nevertheless, it may happen that these roots may be
transformed into one real and two complex conjugate roots. This occurs if $%
\Delta <0$ in Eq. (\ref{5.38a}) and if 
\begin{equation}
\frac{R}{R_{c}}<-\frac{\left( -D_{T}k^{2}+\sigma _{3}k^{2}-\frac{\Omega
^{2}K_{I}k^{4}}{\rho _{0}\sigma _{3}k^{2}}\allowbreak \right) ^{2}}{%
4D_{T}k^{2}\sigma _{3}k^{2}}.  \label{5.38t}
\end{equation}

If we consider the orders of magnitude of the involved quantities $%
D_{T}k^{2}\sim 10^{7}$, $\sigma _{3}k^{2}\sim 10^{8}$ and $\frac{\Omega
^{2}K_{I}k^{4}}{\rho _{0}}\sim 10^{14}$, then in Eq. (\ref{5.38t}) $%
R/R_{c}\lesssim -10^{1}$ is negative. Thereby, Eq. (\ref{5.38t}) implies
that there are two visco-heat propagating modes when $\frac{R}{R_{c}}<0$ and 
$R/R_{c}\lesssim -10^{1}$. According to Eq. (\ref{5.38l}), this occurs if $%
\alpha $ changes its sign and increases by several orders of magnitude,
situation that may be achieved by reversing the direction in which the
temperature gradient is applied, i. e., when heating from below, and by
increasing its intensity. To our knowledge, there are no theoretical
analysis nor exprimental evidence for the existence of visco-heat
propagating modes in nematic liquid crystals under the presence of a
temperature gradient and an uniform gravitational field. Given that in
simple fluids, under these conditions, there are analytical \cite%
{lekkerkerker-boon} and experimental \cite{Boon-Allain-Lallerman} studies
that supports the presence of visco-heat propagation modes, this prediction
suggests that it may be worth to design experiments to corroborate this
phenomenon in nematics.

\subsection{Transverse modes}

The roots of the quadratic polynomial $p_{T}(\lambda )$ of the matrix $N^{T}$
given by Eq. (\ref{5.33e}), are the nematic transverse modes. According to
Eqs. (\ref{5.33bd}), (\ref{5.33c}), (\ref{5.33h}), Eq. (\ref{5.33b}) may be
written as%
\begin{equation}
\frac{\partial }{\partial t}\overrightarrow{Z}^{T}(\overrightarrow{k}%
,t)=-N^{T}\overrightarrow{Z}^{T}(\overrightarrow{k},t)+\overrightarrow{\Xi }%
^{T}(\overrightarrow{k},t),  \label{5.41a}
\end{equation}%
which it is the linear stochastic equation for the transverse variables.

\subsubsection{Shear and director transverse modes}

According to Eq. (\ref{5.33e}) the shear and director transverse modes are
the roots of 
\begin{equation}
\lambda ^{2}-\left( \sigma _{4}k^{2}+\frac{K_{II}k^{2}}{\gamma _{1}}\right)
\lambda +\sigma _{4}k^{2}\frac{K_{II}k^{2}}{\gamma _{1}}+\frac{\lambda
_{+}^{2}K_{II}k^{2}k_{z}^{2}}{\rho _{0}}=0.  \label{5.41c}
\end{equation}%
Following again the approximate method of small quantities used previously,
the small quantities quantities $\sigma _{4}k^{2}$, $K_{II}k^{2}/\gamma _{1}$
and $\lambda _{+}^{2}K_{II}k^{2}k_{z}^{2}/\rho _{0}$, may be identifid in
Eq. (\ref{5.41c}). We define the small or reduced quantities%
\begin{equation}
a_{4}\equiv \frac{\sigma _{4}k^{2}}{\omega },\text{ \ }a_{5}^{\prime }\equiv 
\frac{K_{II}k^{2}}{\gamma _{1}\omega },\text{ \ \ }a_{6}^{\prime }\equiv 
\frac{\lambda _{+}^{2}K_{II}}{\rho _{0}\omega }k^{2}k_{z}^{2}.  \label{5.41d}
\end{equation}%
Since for typical nematics $\lambda _{1}$ is of the order of unity, that $%
\gamma _{1}\sim 10^{-1}$, $\sigma _{4}\sim 10^{-2}$, $K_{II}\sim 10^{-6}$ 
\cite{de gennes}, and also taking into account that $c_{s}\sim 10^{5},$ $%
k\sim 10^{5}$, $g\sim 10^{3}$, the quantities in Eq. (\ref{5.41d}) have the
orders of magnitude $a_{4}\sim 10^{-2}$, $a_{5}^{\prime }\sim 10^{-5}$ and $%
a_{6}^{\prime }\sim 10^{4}$. Therefore, in terms of the reduced variable $%
s\equiv \lambda /\omega $, Eq. (\ref{5.41c}) takes the form%
\begin{equation}
s^{2}+A^{\prime \prime }s+B^{\prime \prime }=0,  \label{5.41f}
\end{equation}%
with%
\begin{equation}
A^{\prime \prime }\equiv -a_{4}-a_{5}^{\prime },  \label{5.41g}
\end{equation}%
\begin{equation}
B^{\prime \prime }\equiv a_{4}a_{5}^{\prime }+\frac{a_{6}^{\prime }}{\omega }%
.  \label{5.41h}
\end{equation}%
The analytic solutions of Eq. (\ref{5.41f}) are%
\begin{equation}
s_{+}=-\frac{1}{2}A^{\prime \prime }+\frac{1}{2}\sqrt{\Delta ^{\prime \prime
}},  \label{5.41i}
\end{equation}%
\begin{equation}
s_{-}=-\frac{1}{2}A^{\prime \prime }-\frac{1}{2}\sqrt{\Delta ^{\prime \prime
}},  \label{5.41j}
\end{equation}%
in which the discriminant is given by%
\begin{equation}
\Delta ^{\prime \prime }\equiv A^{\prime \prime 2}-4B^{\prime \prime }.
\label{5.41k}
\end{equation}%
According to the orders of magnitude of the quantities (\ref{5.41d}), the
discriminant Eq. (\ref{5.41k}) may be simplified to $\Delta ^{\prime }\simeq
a_{4}^{2}-2a_{4}a_{5}^{\prime }>0$, which implies that $\Delta ^{\prime
\prime }>0$ always. Consequently, the solutions (\ref{5.41i}) and (\ref%
{5.41j}) will be real and different, namely,%
\begin{equation}
s_{+}\simeq a_{4}-\frac{a_{6}^{\prime }}{\omega a_{4}},  \label{5.41l}
\end{equation}%
\begin{equation}
s_{-}\simeq a_{5}^{\prime }+\frac{a_{6}^{\prime }}{\omega a_{4}}.
\label{5.41m}
\end{equation}%
As before, they are rewritten as%
\begin{equation}
\lambda _{6}=\sigma _{4}k^{2}-\frac{\lambda _{+}^{2}K_{II}k^{2}k_{z}^{2}}{%
\rho _{0}\sigma _{4}k^{2}},  \label{5.41n}
\end{equation}%
\begin{equation}
\lambda _{7}=\frac{K_{II}k^{2}}{\gamma _{1}}+\frac{\lambda
_{+}^{2}K_{II}k^{2}k_{z}^{2}}{\rho _{0}\sigma _{4}k^{2}}.  \label{5.41o}
\end{equation}%
It should be noted that the shear and director diffusive transverse modes
found previously, Eqs. (\ref{5.41n}) and (\ref{5.41o}), completely match
with those already reported for nematic systems \cite{buka}, \cite{Rusos}, 
\cite{de gennes}.

\section{Discussion and conclusions}

The theoretical results obtained in this work indicate that the presence of
a thermal gradient $\alpha $ and gravitational field $g$ produced its most
significant effect only on the visco-heat $\lambda _{3,4}$\ (formed by the
coupling of the shear an thermal modes) and director $\lambda _{5}$\
longitudinal modes. In these modes the effect is of the order of $10^{-9}$.
In contrast, in the other two remaining sound propagating longitudinal
modes, $\lambda _{1}$ and $\lambda _{2}$, $g$ is the only external force
that produces a small influence of the order of $10^{-24}$. In contrast, the
shear $\lambda _{6}$ and director $\lambda _{7}$ transverse modes are not
affected by these external forces.

The analytical expressions found for the nematodynamic modes are more
general than the previously reported in literature, but when $R=0$, they
reduce to the corresponding expressions already reported for a nematic in
the equilibrium state. Also, in the isotropic limit, these modes reduced to
those of a simple fluid.

When $R$ reaches its critical value $R_{c}$, $R=R_{c}$, $\lambda _{3}$
vanishes, $\lambda _{5}$ is virtually unchanged, while $\lambda _{4}$ has
contributions from the thermal and shear difusive modes. It should be
remaked that this behavior also occurs for a simple fluid.In this case there
are two diffusive modes, one of them also vanishes, and the other one has
contributions from the shear and thermal modes \cite{ortiz}, \cite%
{lekkerkerker-boon}.\ For a simple fluid, these features have been predicted
theoretically, and even more, verified experimentally. These results suggest
that it might be feasible to verify them experimentally for nematics as
well. Our results obtained in this limit do not coincide with those reported
for a $NLC$ \cite{miyakawa}, \cite{migranov}, according to which the
director mode tends to zero, the shear mode does not change and there is an
additional mode which is the sum of the thermal and director modes .

If we consider the orders of magnitude of the involved quantities $%
D_{T}k^{2}\sim 10^{7}$, $\sigma _{3}k^{2}\sim 10^{8}$ and $\frac{\Omega
^{2}K_{I}k^{4}}{\rho _{0}}\sim 10^{14}$, then from Eq. (\ref{5.38t}) $%
R/R_{c}\lesssim -10^{1}$. Thereby, Eq. (\ref{5.38t}) implies that there are
two visco-heat \textit{propagating} modes when $\frac{R}{R_{c}}<0$ and $%
R/R_{c}\lesssim -10^{1}$, a prediction which is not contained in Refs. \cite%
{miyakawa}, \cite{migranov}, and is valid for a simple fluid \cite%
{lekkerkerker-boon}, \cite{Boon-Allain-Lallerman}. Since the existence of
these propagative modes has only been predicted and verified experimentally
in simple fluids, our prediction for $NLC$ suggests that their existence
might be also verified experimentally.

In the literature, the nematic longitudinal hydrodynamic modes in a steady
state have been studied in Refs. \cite{miyakawa}, \cite{migranov} for the
same $NESS$ considered in this work. These works predict that the thermal
and director diffusive modes are coupled. We believe that this result is not
correct, because in the isotropic limit, these modes do not reduce to the
corresponding visco-heat modes of a simple fluid \cite{lekkerkerker-boon}.
In contrast, the analytical expressions that we have found for these
nematodynamic modes imply that the heat and shear modes of the $NLC$ are
coupled and do reduce to those of simple fluid in the isotropic limit.

\newpage

\section*{Appendix A}

The sums of stochastic noises $\zeta _{m}$ (whith $m=1\ldots 6$) in Eqs. (%
\ref{5.33g}) and (\ref{5.33h}) are defined as%
\begin{equation}
\zeta _{1}\left( \overrightarrow{k},t\right) \equiv -i\left( \frac{\gamma -1%
}{\rho _{0}T_{0}c_{p}}\right) ^{1/2}k_{j}\widetilde{\pi }_{j},
\label{5.33ha}
\end{equation}%
\begin{equation}
\zeta _{2}\left( \overrightarrow{k},t\right) \equiv -\frac{k_{j}}{\rho
_{0}^{1/2}k}\left( k_{x}\widetilde{\Sigma }_{xj}+k_{y}\widetilde{\Sigma }%
_{yj}+k_{z}\widetilde{\Sigma }_{zj}\right) ,  \label{5.33hb}
\end{equation}%
\begin{equation}
\zeta _{3}\left( \overrightarrow{k},t\right) \equiv -i\left( \frac{1}{\rho
_{0}T_{0}c_{p}}\right) ^{1/2}k_{j}\widetilde{\pi }_{j},  \label{5.33hc}
\end{equation}%
\begin{equation}
\zeta _{4}\left( \overrightarrow{k},t\right) \equiv -\frac{ik_{z}}{\rho
_{0}^{1/2}}\frac{k_{j}}{k^{2}}\left( k_{x}\widetilde{\Sigma }_{xj}+k_{y}%
\widetilde{\Sigma }_{yj}+k_{z}\widetilde{\Sigma }_{zj}\right) +\frac{ik_{j}}{%
\rho _{0}^{1/2}}\widetilde{\Sigma }_{zj},  \label{5.33hd}
\end{equation}%
\begin{equation}
\zeta _{5}\left( \overrightarrow{k},t\right) \equiv i\rho _{0}^{1/2}c_{s}%
\frac{k_{x}}{k}\widetilde{\Upsilon }_{x}+i\rho _{0}^{1/2}c_{s}\frac{k_{y}}{k}%
\widetilde{\Upsilon }_{y},  \label{5.33he}
\end{equation}%
\begin{equation}
\zeta _{6}\left( \overrightarrow{k},t\right) \equiv -\frac{k_{j}}{\rho
_{0}^{1/2}k}\left( k_{x}\widetilde{\Sigma }_{yj}-k_{y}\widetilde{\Sigma }%
_{xj}\right) ,  \label{5.33hf}
\end{equation}%
\begin{equation}
\zeta _{7}\left( \overrightarrow{k},t\right) \equiv i\rho _{0}^{1/2}c_{s}%
\frac{k_{x}}{k}\widetilde{\Upsilon }_{y}-i\rho _{0}^{1/2}c_{s}\frac{k_{y}}{k}%
\widetilde{\Upsilon }_{x},  \label{5.33hg}
\end{equation}%
where $j=x,y,z.$

The autocorrelations and cross-correlations of the stochastic noises at the
two different points $\left( \overrightarrow{k},\omega \right) $ and $\left( 
\overrightarrow{q},w\right) $, are calculated by using the
fluctuation-dissipation relations Eqs. (\ref{3-28})-(\ref{3-30}) averaged
over the steady state. They are given by

\begin{gather}
\left\langle \zeta _{1}\left( \overrightarrow{k},\omega \right) \zeta
_{1}^{\ast }\left( \overrightarrow{q},w\right) \right\rangle ^{st}=\frac{%
2k_{B}\widetilde{T}^{st}\left( \overrightarrow{k},\overrightarrow{q},%
\overrightarrow{s}\right) (\gamma -1)}{\rho _{0}c_{p}}\left[ \kappa _{\perp
}\left( k_{x}q_{x}+k_{y}q_{y}\right) \right.  \notag \\
\left. +\kappa _{\parallel }k_{z}q_{z}\delta \left( \omega -w\right) \right]
,  \label{5.33ia}
\end{gather}

\begin{gather}
\left\langle \zeta _{2}\left( \overrightarrow{k},\omega \right) \zeta
_{2}^{\ast }\left( \overrightarrow{q},w\right) \right\rangle ^{st}=\frac{%
2k_{B}\widetilde{T}^{st}\left( \overrightarrow{k},\overrightarrow{q},%
\overrightarrow{s}\right) }{\rho _{0}kq}[\left( \nu _{2}+\nu _{4}\right)
\left( k_{x}^{2}q_{x}^{2}+k_{y}^{2}q_{y}^{2}\right)  \notag \\
+\left( \nu _{4}-\nu _{2}\right) \left(
k_{y}^{2}q_{x}^{2}+k_{x}^{2}q_{y}^{2}\right) +4\nu
_{2}k_{x}k_{y}q_{y}q_{x}+4\nu _{3}\left( k_{x}q_{x}+k_{y}q_{y}\right)
k_{z}q_{z}  \notag \\
+\nu _{5}\left( q_{x}^{2}+q_{y}^{2}\right) k_{z}^{2}+\nu _{5}\left(
k_{x}^{2}+k_{y}^{2}\right) q_{z}^{2}  \notag \\
+\left( 2\nu _{1}+\nu _{2}-\nu _{4}+2\nu _{5}\right)
k_{z}^{2}q_{z}^{2}]\delta \left( \omega -w\right) ,  \label{5.33ib}
\end{gather}

\begin{gather}
\left\langle \zeta _{3}\left( \overrightarrow{k},\omega \right) \zeta
_{3}^{\ast }\left( \overrightarrow{q},w\right) \right\rangle ^{st}=\frac{%
2k_{B}\widetilde{T}^{st}\left( \overrightarrow{k},\overrightarrow{q},%
\overrightarrow{s}\right) }{\rho _{0}c_{p}}\left[ \kappa _{\perp }\left(
k_{x}q_{x}+k_{y}q_{y}\right) \right.  \notag \\
\left. +\kappa _{\parallel }k_{z}q_{z}\right] \delta \left( \omega -w\right)
,  \label{5.33ic}
\end{gather}

\begin{gather}
\left\langle \zeta _{4}\left( \overrightarrow{k},\omega \right) \zeta
_{4}^{\ast }\left( \overrightarrow{q},w\right) \right\rangle ^{st}=\frac{%
2k_{B}\widetilde{T}^{st}\left( \overrightarrow{k},\overrightarrow{q},%
\overrightarrow{s}\right) k_{z}q_{z}}{\rho _{0}k^{2}q^{2}}\left[ \left( \nu
_{2}+\nu _{4}\right) \left( k_{x}^{2}q_{x}^{2}+k_{y}^{2}q_{y}^{2}\right)
\right.  \notag \\
\left. +\left( \nu _{4}-\nu _{2}\right) \left(
k_{y}^{2}q_{x}^{2}+k_{x}^{2}q_{y}^{2}\right) \right. \left. +4\nu
_{2}k_{x}k_{y}q_{y}q_{x}+4\nu _{3}\left( k_{x}q_{x}+k_{y}q_{y}\right)
k_{z}q_{z}\right.  \notag \\
\left. +\nu _{5}\left( q_{x}^{2}+q_{y}^{2}\right) k_{z}^{2}+\nu _{5}\left(
k_{x}^{2}+k_{y}^{2}\right) q_{z}^{2}\right.  \notag \\
\left. +\left( 2\nu _{1}+\nu _{2}-\nu _{4}+2\nu _{5}\right)
k_{z}^{2}q_{z}^{2}]\delta \left( \omega -w\right) \right] ,  \label{5.33id}
\end{gather}

\begin{equation}
\left\langle \zeta _{5}\left( \overrightarrow{k},\omega \right) \zeta
_{5}^{\ast }\left( \overrightarrow{q},w\right) \right\rangle ^{st}=\frac{%
2k_{B}\widetilde{T}^{st}\left( \overrightarrow{k},\overrightarrow{q},%
\overrightarrow{s}\right) \rho _{0}c_{s}^{2}}{\gamma _{1}kq}\left(
k_{x}q_{x}+k_{y}q_{y}\right) \delta \left( \omega -w\right) ,  \label{5.33ie}
\end{equation}

\begin{gather}
\left\langle \zeta _{6}\left( \overrightarrow{k},\omega \right) \zeta
_{6}^{\ast }\left( \overrightarrow{q},w\right) \right\rangle ^{st}=\frac{%
2k_{B}\widetilde{T}^{st}\left( \overrightarrow{k},\overrightarrow{q},%
\overrightarrow{s}\right) }{\rho _{0}kq}\left[ \nu _{2}\left(
k_{x}q_{x}+k_{y}q_{y}\right) ^{2}\right.  \notag \\
\left. -\nu _{2}\left( k_{y}q_{x}-k_{x}q_{y}\right) ^{2}+\nu _{3}\left(
k_{x}q_{x}+k_{y}q_{y}\right) k_{z}q_{z}\right] \delta \left( \omega
-w\right) ,  \label{5.33if}
\end{gather}

\begin{equation}
\left\langle \zeta _{7}\left( \overrightarrow{k},\omega \right) \zeta
_{7}^{\ast }\left( \overrightarrow{q},w\right) \right\rangle ^{st}=\frac{%
2k_{B}\widetilde{T}^{st}\left( \overrightarrow{k},\overrightarrow{q},%
\overrightarrow{s}\right) \rho _{0}c_{s}^{2}}{\gamma _{1}kq}\left(
k_{x}q_{x}+k_{y}q_{y}\right) \delta \left( \omega -w\right) ;  \label{5.33ig}
\end{equation}%
and by%
\begin{gather}
\left\langle \zeta _{1}\left( \overrightarrow{k},\omega \right) \zeta
_{3}^{\ast }\left( \overrightarrow{q},w\right) \right\rangle ^{st}=\frac{%
2k_{B}\widetilde{T}^{st}\left( \overrightarrow{k},\overrightarrow{q},%
\overrightarrow{s}\right) \left( \gamma -1\right) ^{1/2}}{\rho _{0}c_{p}}
\label{5.33ih} \\
\left[ \kappa _{\perp }\left( k_{x}q_{x}+k_{y}q_{y}\right) +\kappa
_{\parallel }k_{z}q_{z}\right] \delta \left( \omega -w\right) ,  \notag
\end{gather}%
\begin{gather}
\left\langle \zeta _{3}\left( \overrightarrow{k},\omega \right) \zeta
_{1}^{\ast }\left( \overrightarrow{q},w\right) \right\rangle ^{st}=\frac{%
2k_{B}\widetilde{T}^{st}\left( \overrightarrow{k},\overrightarrow{q},%
\overrightarrow{s}\right) \left( \gamma -1\right) ^{1/2}}{\rho _{0}c_{p}}
\label{5.33ii} \\
\left[ \kappa _{\perp }\left( k_{x}q_{x}+k_{y}q_{y}\right) +\kappa
_{\parallel }k_{z}q_{z}\right] \delta \left( \omega -w\right) ,  \notag
\end{gather}

\begin{gather}
\left\langle \zeta _{2}\left( \overrightarrow{k},\omega \right) \zeta
_{4}^{\ast }\left( \overrightarrow{q},w\right) \right\rangle ^{st}=-\frac{%
2ik_{B}\widetilde{T}^{st}\left( \overrightarrow{k},\overrightarrow{q},%
\overrightarrow{s}\right) q_{z}}{\rho _{0}kq^{2}}\left[ \left( \nu _{2}+\nu
_{4}\right) \left( k_{x}^{2}q_{x}^{2}+k_{y}^{2}q_{y}^{2}\right) \right. 
\notag \\
+\left( \nu _{4}-\nu _{2}\right) \left(
k_{y}^{2}q_{x}^{2}+k_{x}^{2}q_{y}^{2}\right) +4\nu
_{2}k_{x}k_{y}q_{y}q_{x}+4\nu _{3}\left( k_{x}q_{x}+k_{y}q_{y}\right)
k_{z}q_{z}  \notag \\
+\nu _{5}\left( q_{x}^{2}+q_{y}^{2}\right) k_{z}^{2}+\nu _{5}\left(
k_{x}^{2}+k_{y}^{2}\right) q_{z}^{2}  \notag \\
+\left( 2\nu _{1}+\nu _{2}-\nu _{4}+2\nu _{5}\right)
k_{z}^{2}q_{z}^{2}]\delta \left( \omega -w\right) ,  \label{5.33ij}
\end{gather}

\begin{gather}
\left\langle \zeta _{4}\left( \overrightarrow{k},\omega \right) \zeta
_{2}^{\ast }\left( \overrightarrow{q},w\right) \right\rangle ^{st}=\frac{%
2ik_{B}\widetilde{T}^{st}\left( \overrightarrow{k},\overrightarrow{q},%
\overrightarrow{s}\right) k_{z}}{\rho _{0}k^{2}q}[\left( \nu _{2}+\nu
_{4}\right) \left( k_{x}^{2}q_{x}^{2}+k_{y}^{2}q_{y}^{2}\right)  \notag \\
+\left( \nu _{4}-\nu _{2}\right) \left(
k_{y}^{2}q_{x}^{2}+k_{x}^{2}q_{y}^{2}\right) +4\nu
_{2}k_{x}k_{y}q_{y}q_{x}+4\nu _{3}\left( k_{x}q_{x}+k_{y}q_{y}\right)
k_{z}q_{z}  \notag \\
+\nu _{5}\left( q_{x}^{2}+q_{y}^{2}\right) k_{z}^{2}+\nu _{5}\left(
k_{x}^{2}+k_{y}^{2}\right) q_{z}^{2}  \notag \\
+\left( 2\nu _{1}+\nu _{2}-\nu _{4}+2\nu _{5}\right)
k_{z}^{2}q_{z}^{2}]\delta \left( \omega -w\right) .  \label{5.33ik}
\end{gather}%
In Eqs. (\ref{5.33ia})-(\ref{5.33ik}), $\widetilde{T}^{st}\left( 
\overrightarrow{k},\overrightarrow{q},\overrightarrow{s}\right) $ can be
identified as the spatial Fourier transform (\ref{5.28a}) of (\ref{2-3}), 
\begin{align}
\widetilde{T}^{st}(\overrightarrow{k},\overrightarrow{q},\overrightarrow{s}%
)& \equiv T_{0}\delta \left( \overrightarrow{k}-\overrightarrow{q}\right)
\label{5.33il} \\
& -\frac{\alpha }{2is}\left[ \delta \left( \overrightarrow{k}-%
\overrightarrow{q}-\overrightarrow{s}\right) -\delta \left( \overrightarrow{k%
}-\overrightarrow{q}+\overrightarrow{s}\right) \right] .  \notag
\end{align}

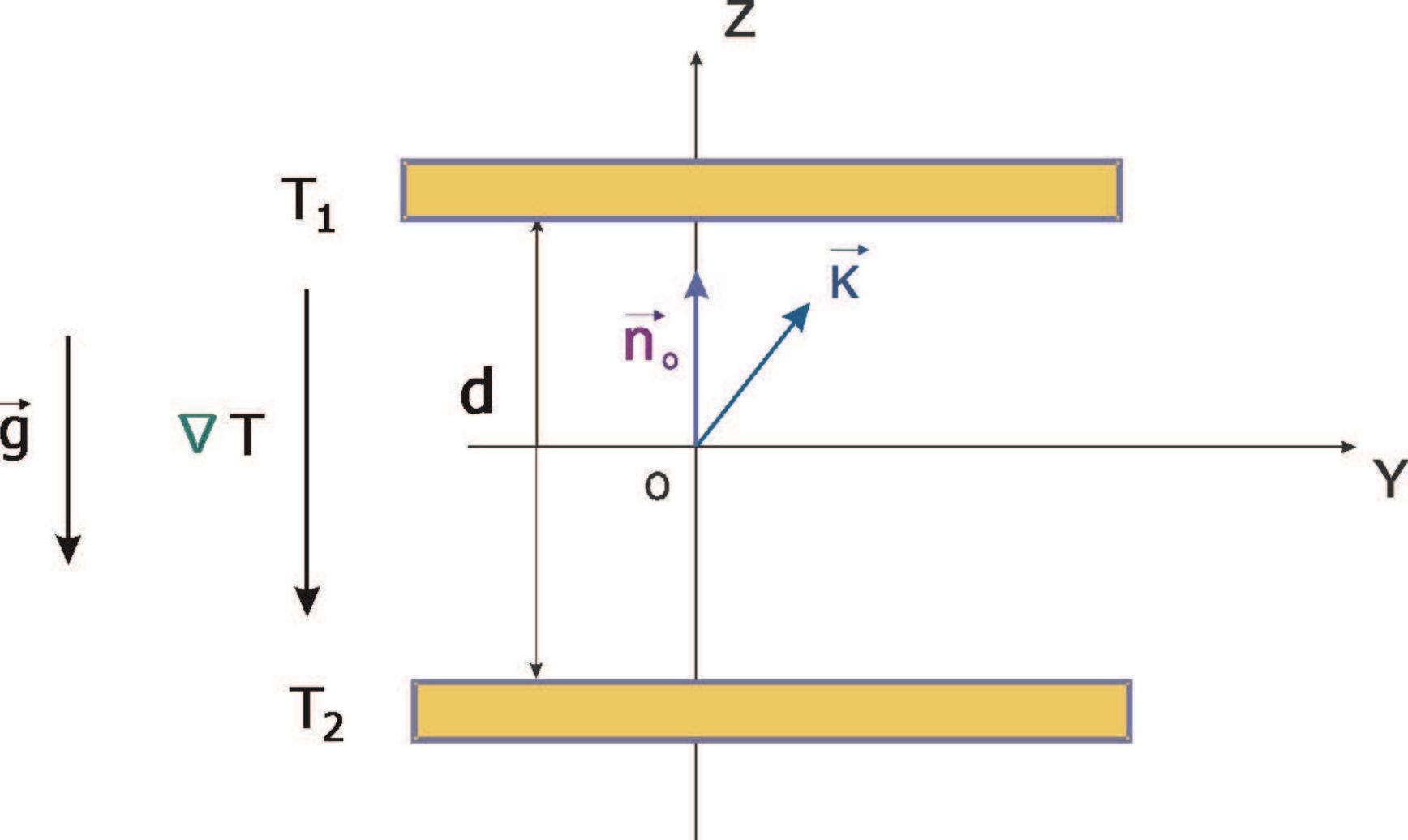
\end{document}